\documentclass[aps,pre,11pt]{revtex4}
\usepackage{amsthm}
\usepackage{newlfont}
\usepackage{graphicx}
\usepackage{subfigure}
\usepackage[section]{placeins}
\usepackage{psfrag}
\usepackage{color}
\usepackage{caption2}
\usepackage{flafter}
\usepackage{bm}%
\usepackage{amsthm}
\usepackage{newlfont}
\usepackage{graphicx}
\usepackage{subeqn}
\usepackage{subfigure}
\usepackage[section]{placeins}
\usepackage{psfrag}
\usepackage{caption2}
\usepackage{flafter}
\usepackage{bm}
\usepackage{bbding}
\usepackage{pifont}
\usepackage{wasysym}
\usepackage{amssymb}
\usepackage{dcolumn}
\usepackage{bm}

\begin{document}
\title{Modulation instability induced supercontinuum generation in liquid core suspended photonic crystal fiber with cubic-quintic nonlinearities}
\author{A. Sharafali$^{1}$, A.K. Shafeeque Ali$^{2}$ and M. Lakshmanan$^{2}$}
 \address{1.Department of Physics, Pondicherry University, Pondicherry-605 014, India.\\ 2.Department of Nonlinear Dynamics, School of Physics, Bharathidasan University, Tiruchirappalli-620 024, India.}
\begin{abstract}
In this paper, we propose a liquid core suspended photonic crystal fiber (LCSPCF)
as a potential waveguide structure for nonlinear applications. We emphasize
the dramatic improvement of the nonlinear properties of the PCF through the infiltration
of liquid and highlight the effect of suspension in the further enhancement of nonlinearity.
To demonstrate the capability of LCSPCF, we perform a modulation
instability (MI) study and illustrate the MI enhancement in the LCSPCF against
conventional PCF. As the LCSPCF possesses high nonlinearity, it is important
to consider the effects of higher-order nonlinearities. We observe that the quintic
nonlinearity (QN) increases the MI for cooperating nonlinearities while it suppresses
the MI for competing nonlinearities. We have also importantly studied the effect
of QN on the MI induced supercontinuum generation (SCG) for silica and $CS_2$
liquid core PCF. Thus, we propose the LCSPCF as a platform for nonlinear applications such as SCG, pulse
compression, etc., owing to the presence of high and controllable nonlinear coefficients.
\end{abstract}
\maketitle
\section{Introduction}
Photonic crystal fiber (PCF) is a special kind of optical waveguide based on the photonic crystals and it has numerous applications in the field of science and engineering \cite{Knight, Knightt, Knighttt}. In contrast to the conventional optical fiber, PCF exhibits unusual characteristics, including  controlled chromatic dispersion \cite{Saitoh}, endlessly single-mode propagation \cite{Birks}, high birefringence \cite{ort} and extreme nonlinearity. Further enhancement of all the associated parameters can be achieved by engineering the geometrical parameters and by the choice of materials of the waveguides. This can be achieved by tailoring  the geometry and size of the cladding air holes \cite{Saitoh,Birks,ort,Chen,Ho,Joannopoulos,Poli,Polii,Liu,Sere,Kibler}. In the literature, one can observe that numerous techniques have been proposed for the enhancement of the linear and nonlinear characteristics of the PCFs. Particularly, the PCFs find a wide interest in nonlinear optics, thanks to the customized design architecture enabling it as an obvious choice. Non-silica technology is one of the possible ways of increasing the nonlinearity of the PCF.

In the present work, we introduce a novel suspended photonic crystal fiber with liquid in the central core of the fiber for nonlinear studies and applications. Even though investigations on liquid core PCF have been going on for some time, the need for new and hybrid materials for different applications is motivating research in optical fiber technology.  For example, index oils \cite{Wang1,Zheng}, polymers \cite{Huang} and  metals \cite{Spittel} can be used to infiltrate into the air holes of the fiber to form non-silica-based PCFs. This kind of PCF is suitable for various applications like switching, communication, optical sensing, etc. Very low chromatic dispersion and low confinement loss in liquid-filled PCFs  can be achieved by varying the size of the air holes and the number of air-hole layers in the cladding, respectively~\cite{Yu,Yu1}. Along with this, liquid filled PCFs have applications in fibered photon-pair generation, temperature sensing, optical wavelength conversion \cite{Cordier,Ayyanar,Methaprian,Monfared}, etc. In this connection, the recent study on suspension core photonic crystal fiber by Ghosh \emph{etal.} shows that such a novel kind of PCF can be used to enhance the nonlinearity by changing the geometry of the  air hole \cite{Ghosh}. Hence in this paper, we propose a novel fiber structure known as liquid core suspended photonic crystal fiber (LCSPCF) to enhance the nonlinear effects due to the increased nonlinear coefficient arising from the liquid infiltration and core suspension.

 Among the different nonlinear phenomena of interest, modulational instability (MI) is treated as a central process, considering its great variety of applications. MI is a ubiquitous phenomenon which has stimulated researchers from different fields for more than three decades \cite{Agrawal1,Agrawal2,Agrawal3,ba1,ba2}.  In the field of  nonlinear optics, MI is considered as an exponential growth of the weak perturbation imposed on a co-propagating strong continuous wave as a result of  the interplay between nonlinear and linear dispersive effects \cite{Benjamin,Karpman,Taniuti,Nithi}. The various applications of MI, like the generation of ultra-short pulse trains with high repetition rate, generation of ultra-broadband spectrum in the name of supercontinuum generation, optical-switching, optical amplification, material absorption and loss compensation, optical sensing, dispersion management \cite{Greer:89,Hasegawa,Demircan,Nithyanandan:14,aksa,aksa1,Zambo:11,Nallusamy,Tai}, etc., have excited the imagination of the research community in many fields.

In the above connection, different parameters like pulse power, pump wavelength, nonlinearity coefficient, etc. can be used to enhance the MI process in optical fibers. Since the MI can be a favorable process to produce ultrashort pulses and supercontinuum, enhancement in the MI gain and spectral width by designing more practically feasible and less economic optical fibers is a progressive research area in optics.  Changing optical parameters to enhance the MI gain spectra is a tedious task in conventional optical fibers. Thanks to the development of PCF structures, it offers a high degree of design flexibility to control the fiber characteristics.  In order to increase the nonlinearity, the core regime has to be fabricated with a medium having high nonlinear refractive index. Wang \emph{etal.} have investigated the effect of pump power and wavelength on the MI by using an $As_2Se_3$ chalcogenide glass PCF. Using the high nonlinearity of $As_2Se_3$ PCF, with suitable pump power and wavelength, they have generated an ultra-broadband MI gain \cite{Wang}. Also, pulses of duration of 490fs have been experimentally produced by accessing the MI regime in a xenon-filled hollow core kagom$\acute{e}$ PCF \cite{Tani}. Along lines similar to the above works on the novel PCF structure, in this paper, we study the MI in the proposed LCSPCF and report the cutting-edge advantage of the proposed model over the conventional fiber structures. In this work, we choose carbon disulphide ($CS_2$) liquid as the core material which has a larger nonlinear refractive index than silica. It also provides a wide optical transmission window from visible to mid IR range. Together with these twin properties of high refractive index material in a very small core PCF, we point out that the nonlinearity coefficient of the fiber gets enhanced and the MI in the LCSPCF has been studied for the first time to the best of our knowledge. Further, we have also studied the MI-induced SCG in silica PCF and CS$_2$ LCSPCF. The MI-induced SCG process is also shown to be a noise-driven process where the modulation of the pulse with exponentially increasing noise causes the pulse to break up and eventually leads to spectral broadening \cite{Hasegawa2}. After analyzing the effect of suspension on the SCG, we focus the study on the effect of quintic nonlinearity on the MI-induced SCG in both the proposed models.

The organization of the paper is as follows: Following the above detailed introduction, in Sec.~\ref{Design} we describe the potential fiber design and the choice of parameters. In Sec.~\ref{MI analysis} we explain the MI analysis analytically, followed by a detailed  discussion on the obtained results in Sec.~\ref{Discussion}. In Sec.~\ref{MISCG} we thoroughly investigate the effect of suspension and quintic nonlinearity on the SCG process. Finally, Sec.~\ref{conclusion} provides a conclusion of the paper with a summary of the results.
\section{Designing Liquid core suspended PCF}
\label{Design}
\begin{figure*}[htb]
\centering
\subfigure[]{\label{normal24}\includegraphics[height=3.5 cm, width=3.5 cm]{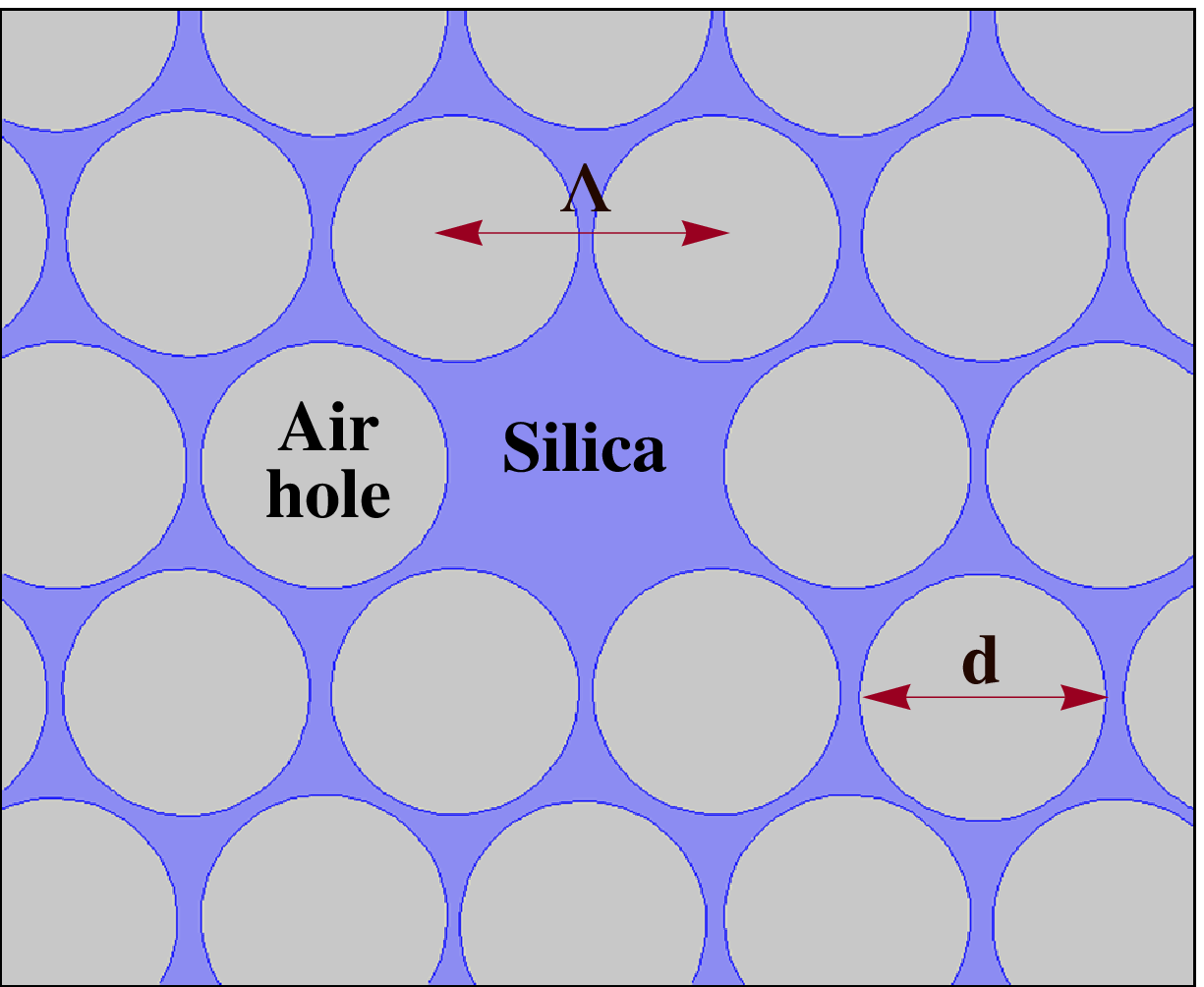}}
\subfigure[]{\label{normal24}\includegraphics[height=3.5 cm, width=3.5 cm]{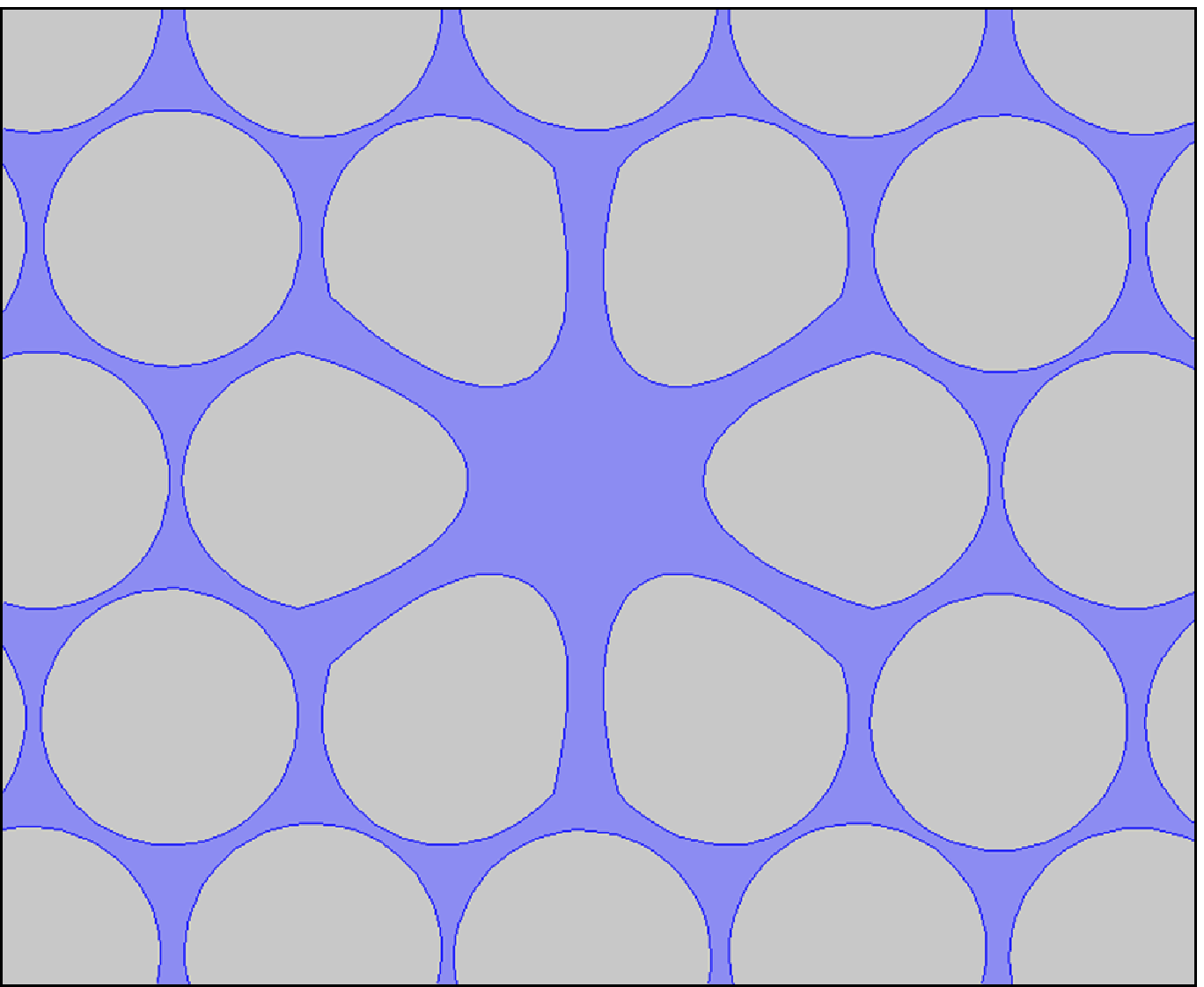}}
\subfigure[]{\label{normal24}\includegraphics[height=3.5 cm, width=3.5 cm]{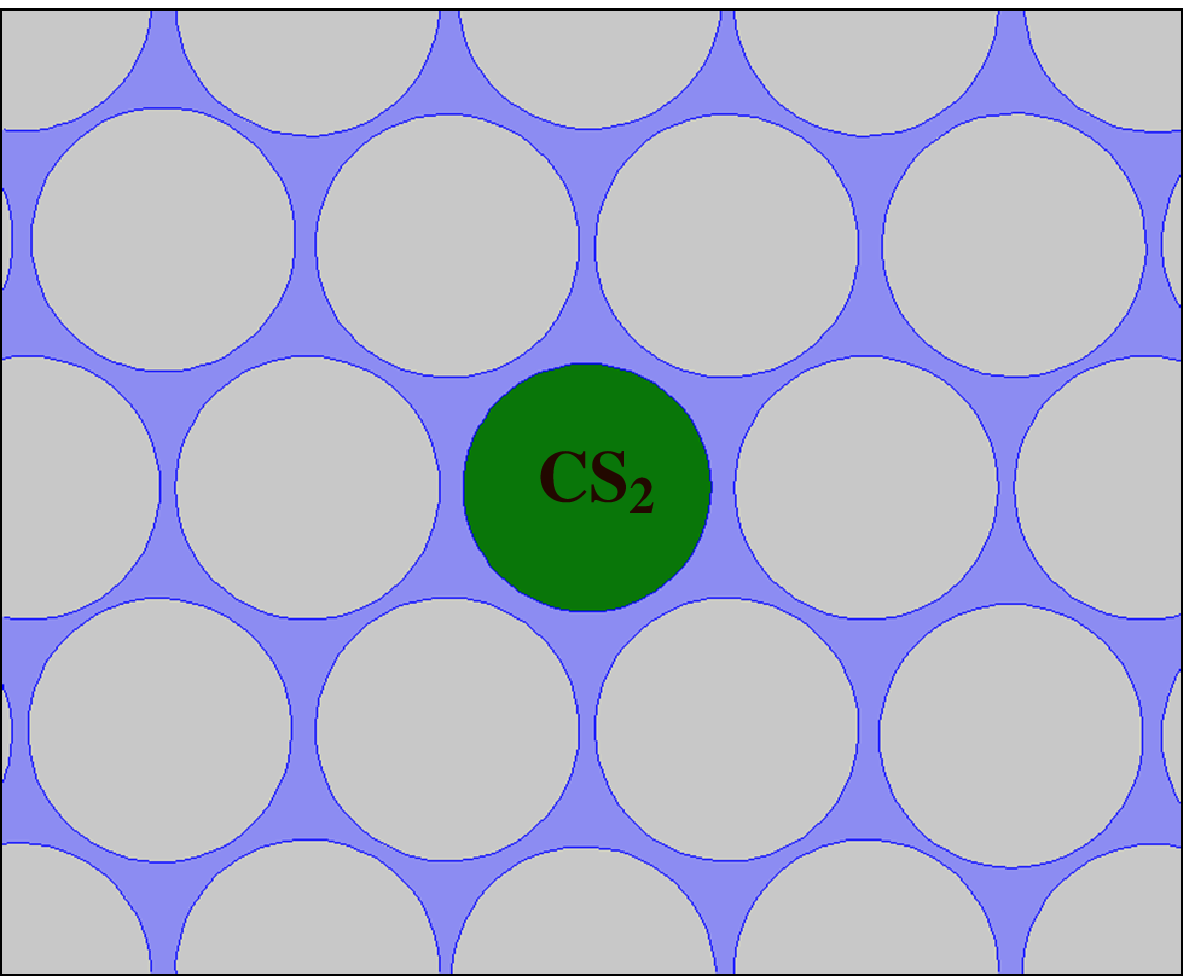}}
\subfigure[]{\label{normal24}\includegraphics[height=3.5 cm, width=3.5 cm]{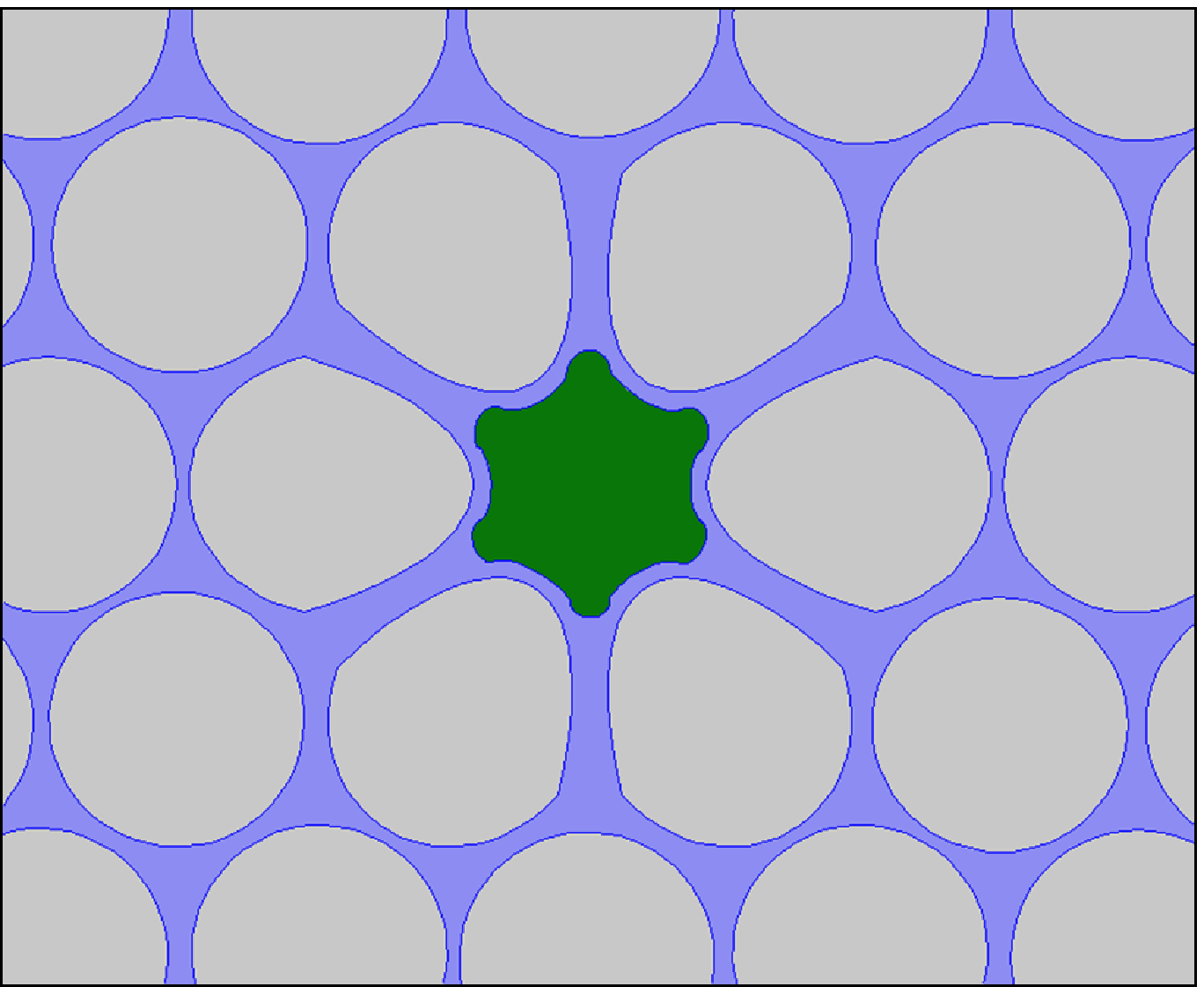}}
\caption{(Color online.) Geometries of the different PCFs proposed: (a) silica core PCF, (b) silica CSPCF, (c) $CS_2$ LCPCF and (d) $CS_2$ LCSPCF}
\label{model}
\end{figure*}

The properties of the PCF are highly dependent on the design and there are numerous PCF structures available in the literature for specific applications. The typical way of controlling the physical parameters is by changing the geometry of the PCF structures such as the pitch (the distance between adjacent air holes, represented by $\Lambda$) and the air hole diameter (denoted by $d$). As the innovation in the design flexibility continuously evolves, there are new means emerging to change the fiber physical parameters beyond just changing the size of the air hole and the pitch to achieve the desired outcome. One of the promising techniques is through a non-silica based technology by the usage of liquids in the fiber core, and such fiber types find very interesting nonlinear applications and they are identified as liquid-core photonic crystal fibers (LCSPCFs).  In this paper, inspired by the recent work of Ghosh \emph{etal}. \cite{Ghosh}, we introduce such liquid core suspended photonic crystal fibers.  Fig. \ref{model} represents the cross sectional view of the design of a conventional PCF and the proposed $CS_2$ filled LCSPCF.  All the proposed PCF structures are modeled by using COMSOL 5.2a software \cite{com} and the mode dynamics is studied using the finite element method (FEM).  Also, we have used the perfect matched layer (PML) boundary condition for modeling all the proposed PCFs. Unlike other absorbing conditions, the PML strongly absorbs the outgoing waves from the computational region without any reflection. It may be noted that a highly efficient circular phase-matched layer has been incorporated to remove the backscattering from the boundaries of the simulation area \cite{Guo}. For better illustration, we have considered the following four possible combinations of the PCF structure as conventional un-suspended  core without and with liquid represented by Figs. \ref{model}(a) and \ref{model}(c) and suspended core without and with liquid as shown in Figs. \ref{model}(b) and \ref{model}(d), respectively.

 In the proposed model, the suspension in the core has been made by deforming the first air hole cladding layer in such a way that it is suspended towards the core regime. Fabrication of the suspended core fiber can be achieved by controlling the fiber drawing parameters such as feed rate, furnace temperature, draw speed and differential pressure in the conventional draw and pull method as demonstrated in Ref.\cite{Ghosh}.  For the simulation, we have used a B$\grave{e}$zier curve to model all the proposed suspended core PCFs. We note that the effective area can appreciably decrease in a solid-core suspended PCF (SCSPCF) as compared to the conventional solid core photonic crystal fiber (SCPCF) without any suspension effect, thus paving the way to elevate the nonlinearity by increasing the suspension factor (SF). It is worth mentioning that the SF cannot be increased beyond a critical value termed $SF_{max}$, which is determined by the PCF parameters $d$ and $\Lambda$.  The value of d/$\Lambda$ = 0.94 with SF values of 1.444 and 1.222 has been typically considered as providing high field confinement in a very small core. For the considered parametric ratio d/$\Lambda$, the lowest value of SF is 0.94 and the critical SF value is 2.5 -(d/$\Lambda$) = 1.56.  Thus the choice of SF value in our problem is within the allowed limit of SF as described in Ref. \cite{Ghosh}.

In the first part of the manuscript, we briefly discuss the effect of fiber suspension and the role of liquid infiltration in the PCF structure under the linear and nonlinear characteristics of the fiber, and compare the results with the conventional SCPCF to facilitate the cutting-edge advantage of the proposed LCSPCF over the conventional counterpart. We have chosen the geometrical parameters of the fiber as d=1.7$\mu m$ and $\Lambda=1.8\mu m$ for both the designs of the PCFs.  The linear refractive indices of the $CS_2$ and silica media (at $20^0$C) are calculated by using the following formulae \cite{GGhosh,Samoc},
\begin{equation}
\label{sieq}
 n_{SiO_2}= \frac{(0.7884+23.583\times10^{-6}T)\times \lambda^2}{\lambda^2-(0.01101+0.5847\times10^{-6} T)}+1.315+6.907
\times10^{-6}T+ \frac{((0.9131+0.5483\times10^{-6}T)\times\lambda^2)}{\lambda^2-100},
\end{equation}
\begin{equation}
 n_{CS_2} (\lambda)=1.580826+1.52389\times10^{-2} \lambda^{-2}+4.8578\times10^{-4} \lambda^{-4}-8.2863\times10^{-5} \lambda^{-6}+1.4619\times10^{-5} \lambda^{-8},\\
\end{equation}
where $\lambda$ is the wavelength in $\mu$m.\\
The dispersion parameters corresponding to the designed fibers are calculated by using the formula,
\begin{eqnarray}
D=-\frac{\lambda}{c}\frac{d^2n_{eff}}{d\lambda^2},
\label{DispersionCoeff}
\end{eqnarray}
where $\lambda$ is the wavelength, $n_{eff}$ is the effective refractive index and $c$ is the velocity of light.
We have used the following expressions for calculating the cubic ($\gamma_1$) and quintic ($\gamma_2$) nonlinearities,
\begin{eqnarray}
\gamma_1=\frac{2\pi n_2}{\lambda A_{eff}},
\label{CubicNonlinearity}
\end{eqnarray}
\begin{eqnarray}
\gamma_2=\frac{2\pi n_4}{\lambda A_{eff}^2}.
\label{QuinticNonlinearity}
\end{eqnarray}
In Eqs. (\ref{CubicNonlinearity}) and (\ref{QuinticNonlinearity}), we have taken the standard values of third-order nonlinear refractive index $(n_2)$ as $3.5\times10^{-20}, 3.1\times10^{-19} m^2/W$  and fifth-order nonlinear refractive index $(n_4)$ as $-3.3\times10^{-36}, 2.2\times10^{-33} m^4/W^2$, respectively, for silica and $CS_2$ \cite{Couris,Besse,Li}. $A_{eff}$ serves as the effective core area of the proposed PCF.
\begin{figure}[htb]
\centering
\subfigure[]{\label{nor1}\includegraphics[height=4.5 cm, width=5 cm]{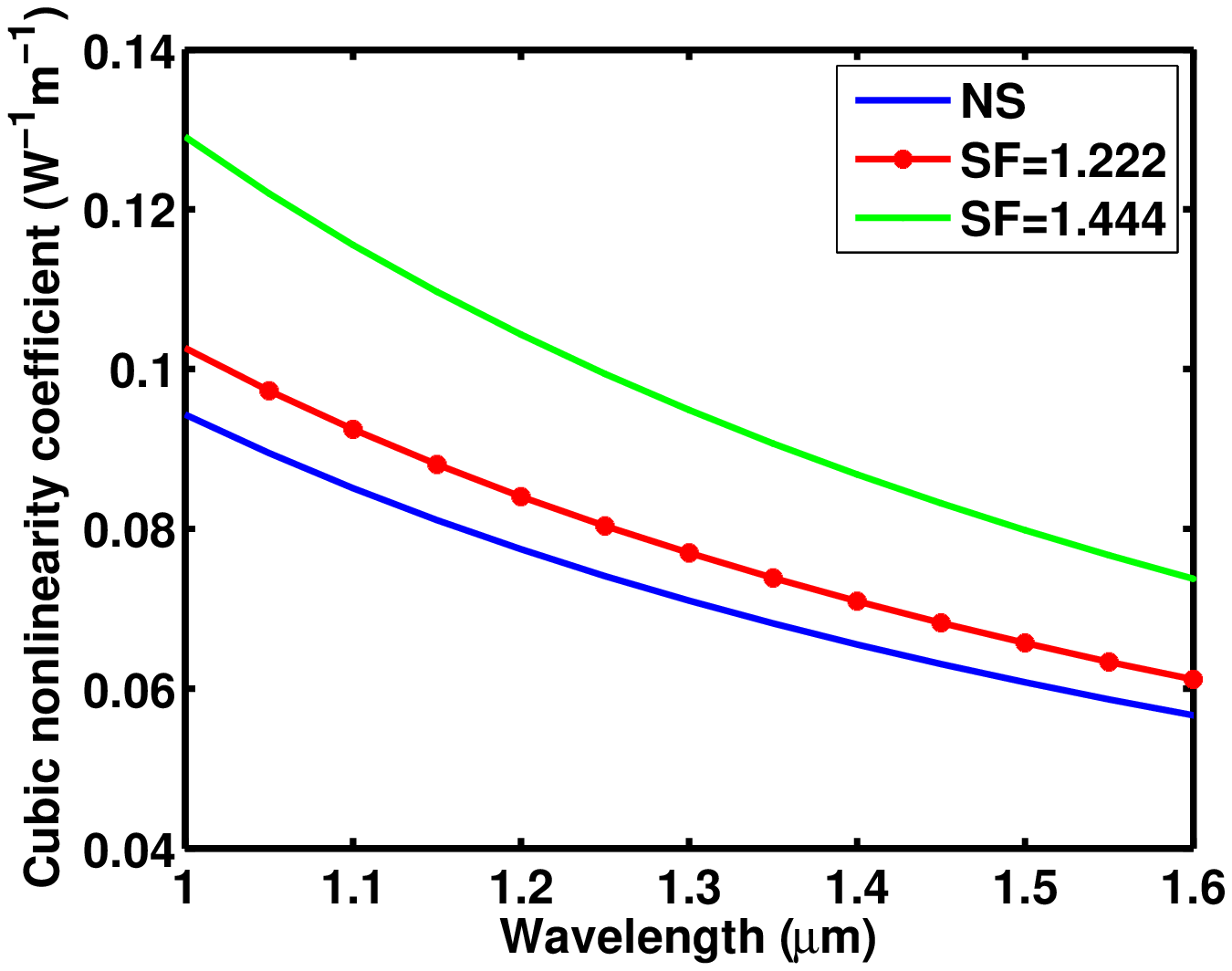}}
\subfigure[]{\label{nor2}\includegraphics[height=4.5 cm, width=5 cm]{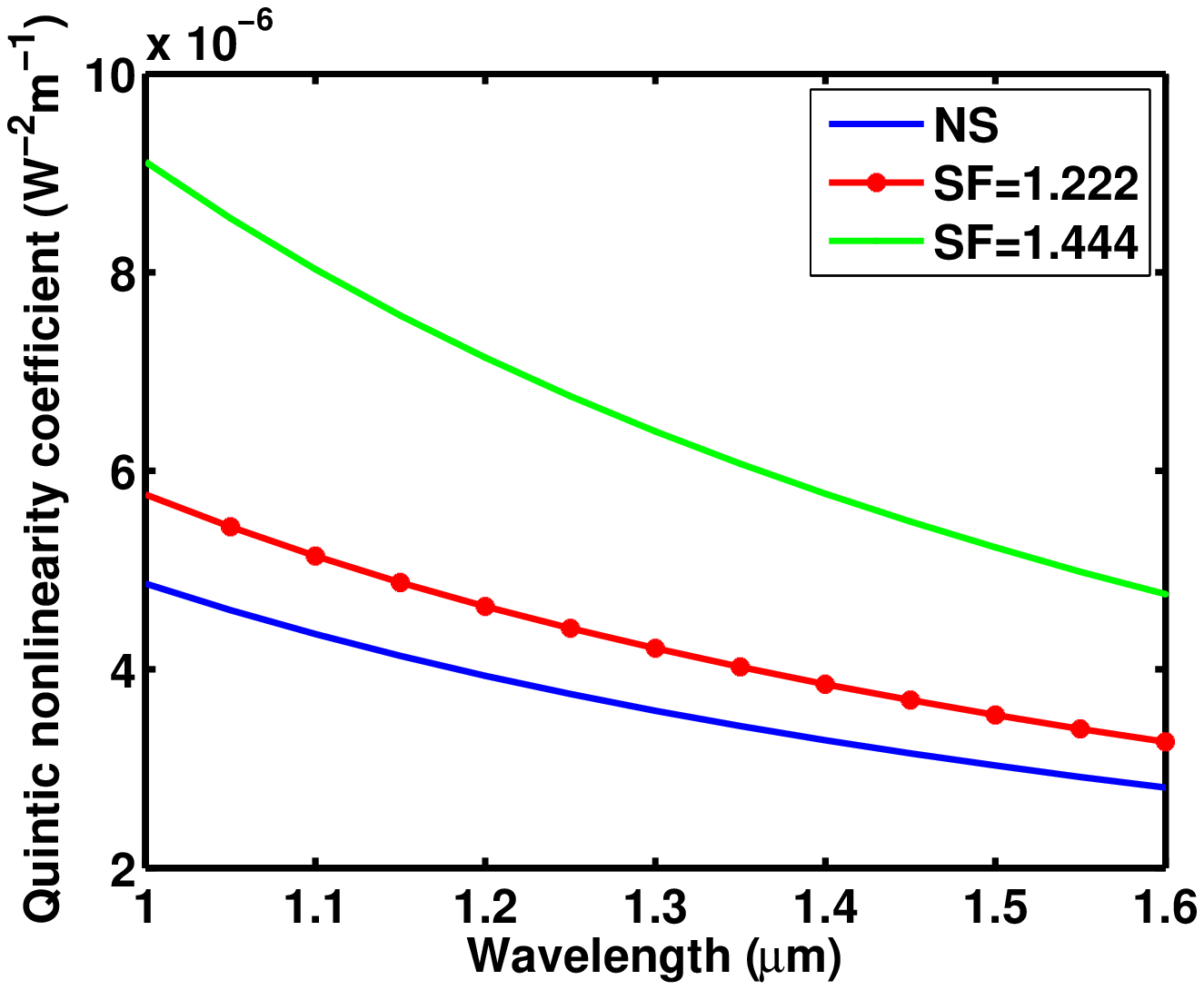}}
\subfigure[]{\label{nor3}\includegraphics[height=4.5 cm, width=5 cm]{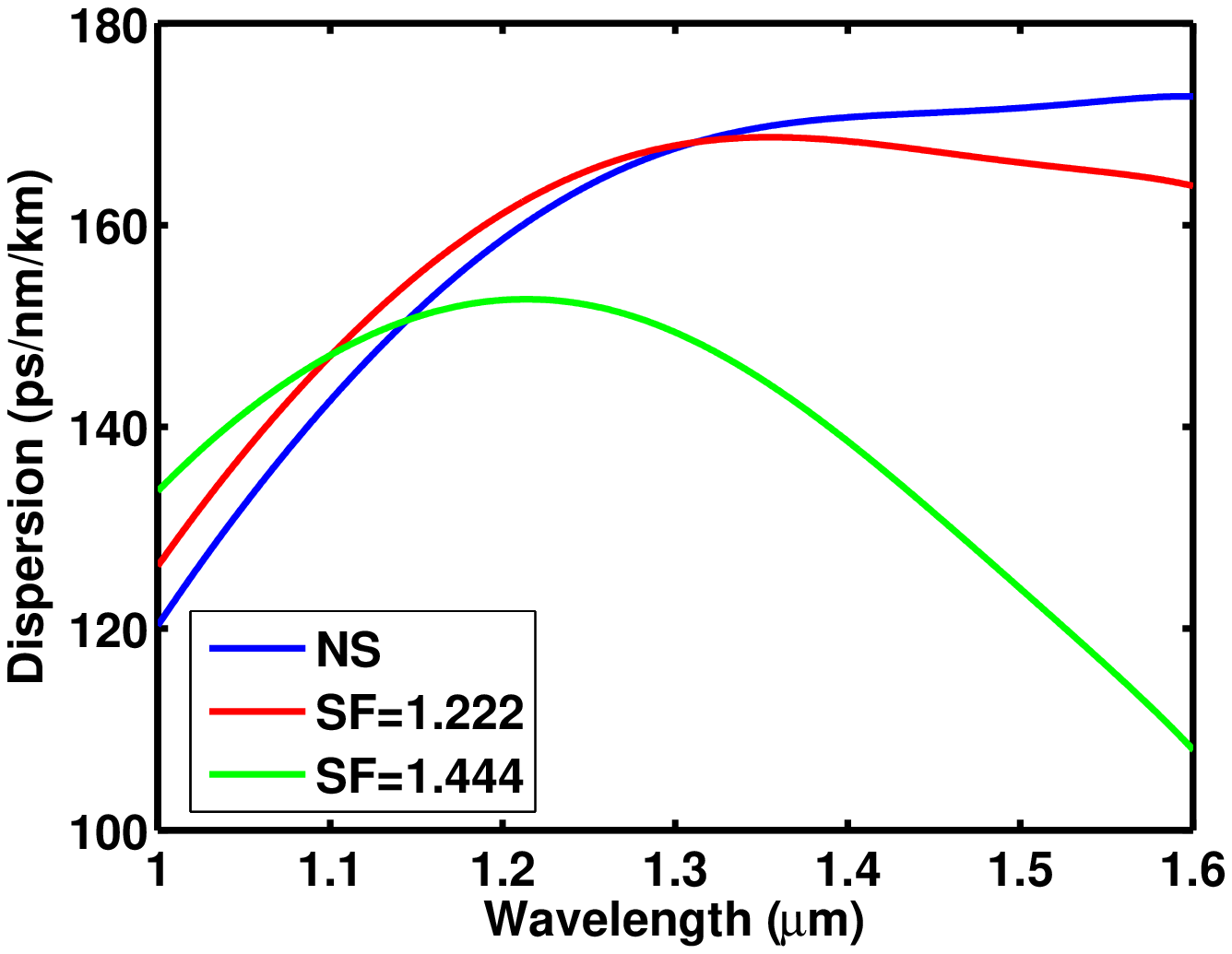}}
\subfigure[]{\label{nor4}\includegraphics[height=4.5 cm, width=5 cm]{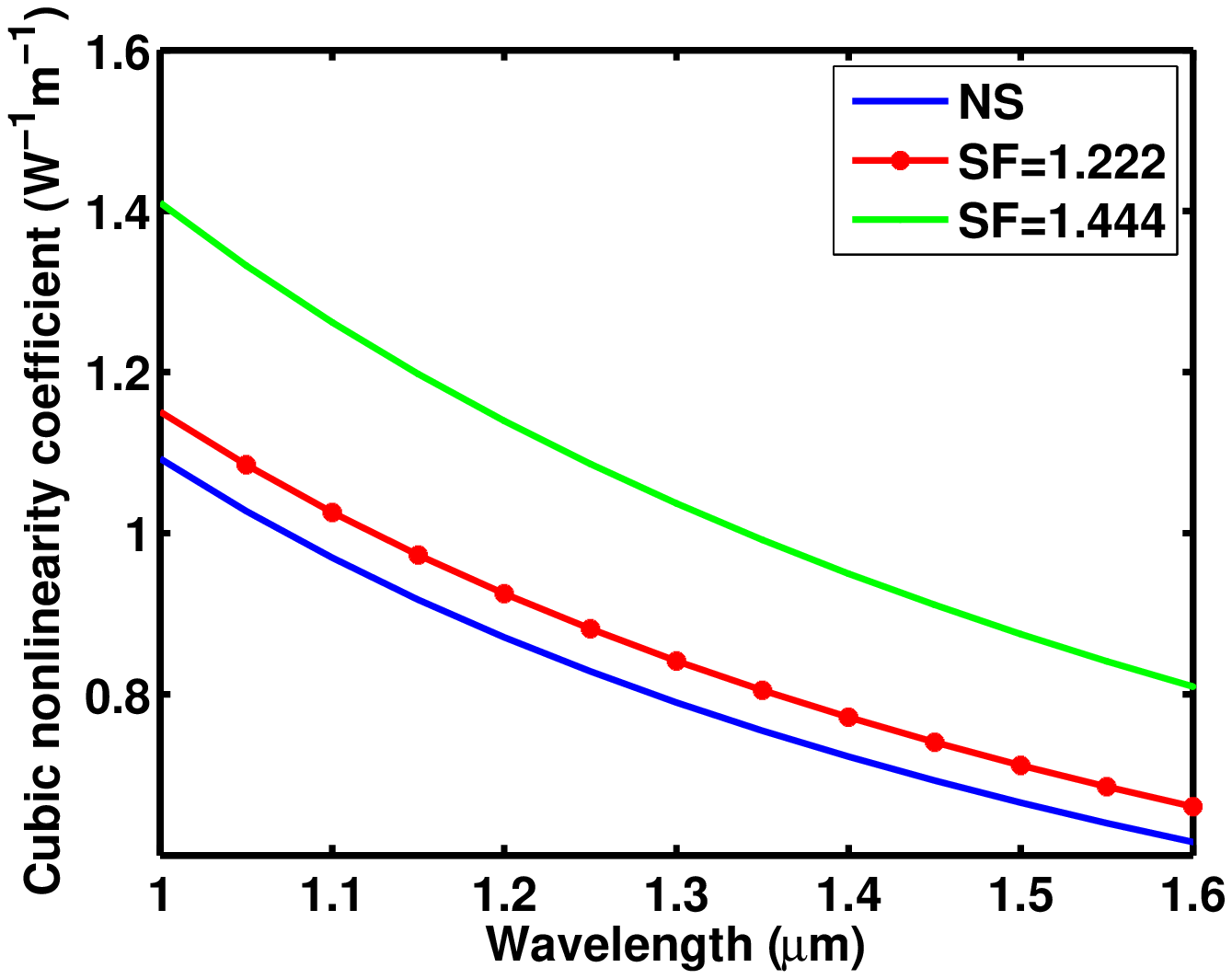}}
\subfigure[]{\label{nor5}\includegraphics[height=4.5 cm, width=5 cm]{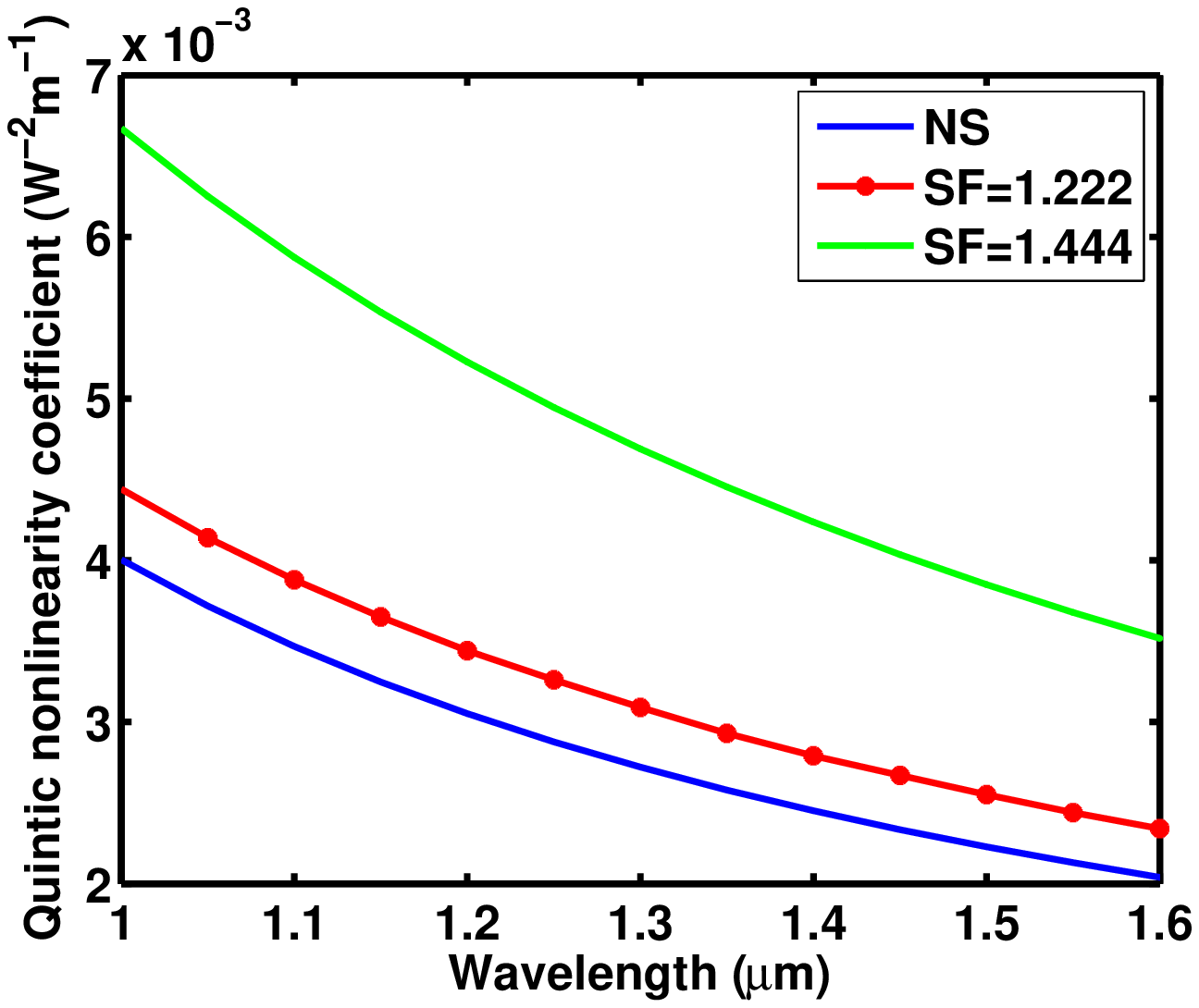}}
\subfigure[]{\label{nor6}\includegraphics[height=4.5 cm, width=5 cm]{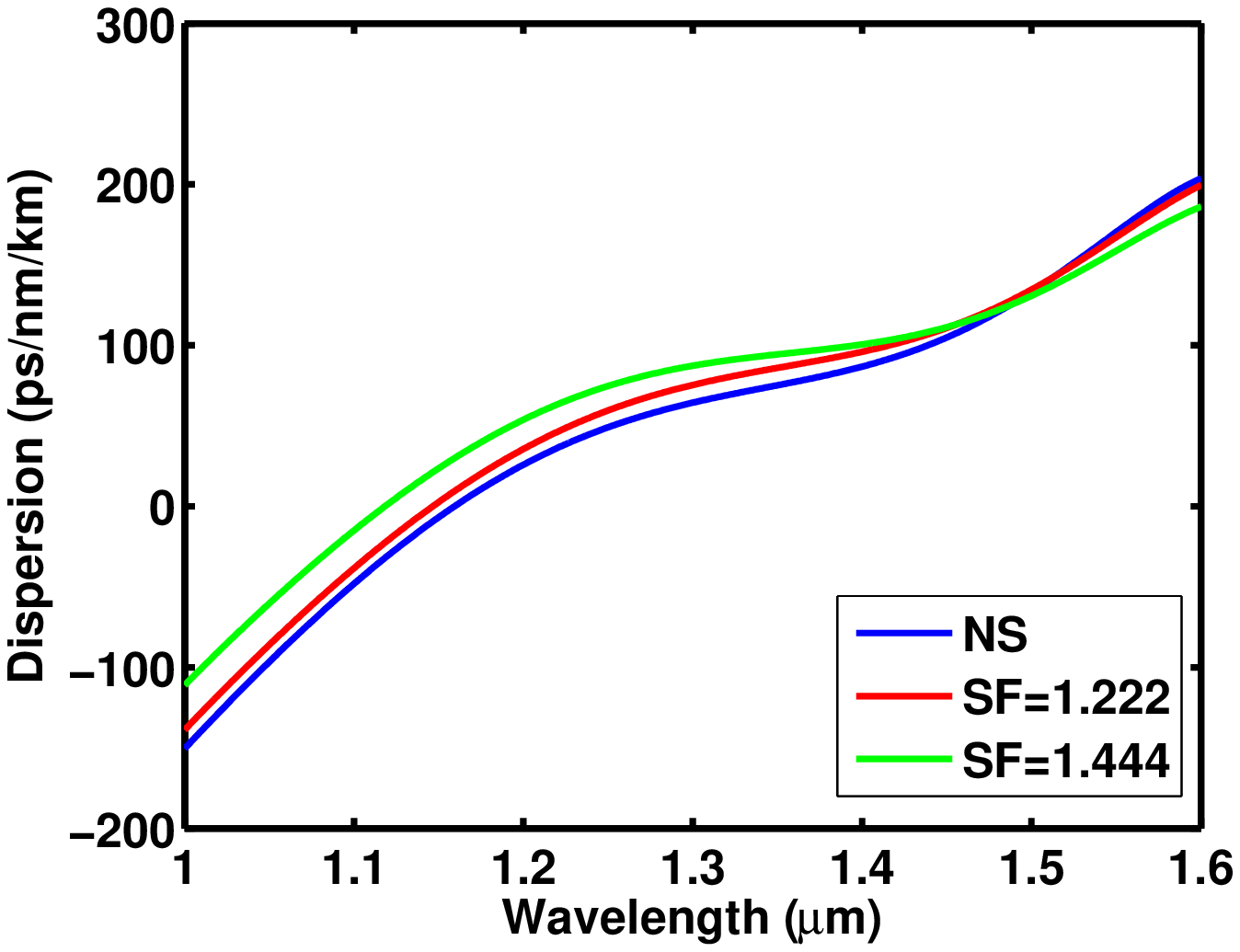}}
\caption{(Color online.) Variation of cubic nonlinearity, quintic nonlinearity and dispersion against wavelength in silica (a, b and c) and CS$_2$ (d, e and f) core PCFs.}
\label{cubic}
\end{figure}
In what follows, we briefly illustrate the effect of liquid infiltration and core suspension in the designed PCF in terms of two of the most significant physical parameters, namely dispersion and nonlinear coefficients. In order to make the discussion self-explanatory, we compare our findings with the conventional PCF and emphasize the merits/demerits of one over the other. Using the expression for the cubic nonlinear coefficient given by Eq. (\ref{CubicNonlinearity}), we estimate the nonlinearity with some representative values of the suspension factor for both the designs of PCFs, namely LCSPCF and SCSPCF. Fig. \ref{cubic} depicts the variation of cubic nonlinearity, quintic nonlinearity and dispersion as a function of wavelength in silica and $CS_2$ core PCFs. The legend NS corresponds to the no-suspension case (core without suspension effect), which enables us to evaluate the role of liquid infiltration in the PCF. It is evident that the incorporation of a nonlinear liquid-like $CS_2$ increases the nonlinearity. For any non-zero value of SF, the nonlinearity inevitably increases regardless of the material in the core. Also, it is certain from Figs. \ref{nor1} and \ref{nor4} that the nonlinearity of the LCSPCF is larger than the silica-based SCPCF due to the large value of the nonlinear refractive index of $CS_2$ liquid, which is further enhanced by the suspension effect. Figs. \ref{nor2} and \ref{nor5} show the variation in the magnitude of the quintic nonlinearity with wavelength, and it is quite straightforward to note that except for the difference in the magnitude between $\gamma_1$ and $\gamma_2$  all the discussion outlined above fairly holds and hence a detailed explanation is avoided in the present work. We also note here that while the quintic nonlinearity present in silica core PCF is competing with the cubic nonlinearity as $\gamma_1>0$ and $\gamma_2<0$ but with $n_4$ being positive in the case of $CS_2$ LCPCF  it is of cooperating in nature as $\gamma_1>0$ and $\gamma_2>0$.

Figs. \ref{nor3} and \ref{nor6} depict the variation of the dispersion parameter as a function of wavelength evaluated by using Eq. (\ref{DispersionCoeff}). Unlike Figs. \ref{nor1} and \ref{nor4}, which show a common trend in the variation of nonlinearity between the two fiber types discussed here, the variation of dispersion is more complex owing to the waveguide contribution to the dispersion. As it is evident from Figs. \ref{nor3} and \ref{nor6}, the variation of the dispersion coefficient is markedly different between the LCSPCF and SCSPCF. Also, we notice that the effect of suspension on dispersion in the LCSPCF is minimal compared to the case of SCSPCF, due to the reduced effective index variation in LCSPCF. We also compute the higher-order dispersion coefficients up to the fourth order from the obtained effective index using the finite difference method. The characteristic values of the nonlinear and dispersion coefficients in the  operating wavelength of 1.55$\mu m$ are tabulated in table \ref{tab1}. From the above discussion it is quite obvious that the nonlinearity effects of the PCF  can be notably enhanced by infiltration of nonlinear liquids and a further increase is possible by suspending the fiber core. From this conclusion we now demonstrate the enhancement of the nonlinear phenomena in the LCSPCF by considering the fundamental process of modulational instability.
\begin{table}
\caption{\label{tab1}Fiber parameters of all proposed PCFs.}
\begin{tabular}{|l|l|l|l|}
\hline
 \multicolumn{4}{l|}{\textbf{\,\,\,\,\,\,\,\,\,\,\,\,\,\,\,\,\,\,\,\,\,\,\,\,\,\,\,\,\,\,\,\,\,\,\,\,\,\,\,\,\,\,\,\,\,\,Silica core}}\\
 \hline
Parameters&NS &SF=1.222 &SF=1.444\\
\hline
$\beta_2 (ps^2/m)$&-0.2200&-0.2109&-0.1484\\
$\beta_3 (ps^3/m)$&4.0335$\times10^{-5}$&3.2829$\times10^{-5}$&1.1369$\times10^{-5}$\\
$\beta_4(ps^4/m)$&-8.3333$\times10^{-9}$&-9.8911$\times10^{-9}$&-4.3628$\times10^{-9}$\\
$\gamma_1 (W^{-1}m^{-1})$&0.0586 &0.0633&0.0767\\
$\gamma_2 (W^{-2}m^{-1})$&-2.91$\times10^{-6}$&-3.40$\times10^{-6}$&-4.98$\times10^{-6}$\\
\hline
\multicolumn{4}{l|}{\textbf{\,\,\,\,\,\,\,\,\,\,\,\,\,\,\,\,\,\,\,\,\,\,\,\,\,\,\,\,\,\,\,\,\,\,\,\,\,\,\, \,\,\,\,\,\,\, CS$_2$ core}}\\
 \hline
Parameters&NS &SF=1.222 &SF=1.444\\
\hline
$\beta_2 (ps^2/m)$&-0.2173&-0.2138&-0.2029\\
$\beta_3 (ps^3/m)$&1.6811$\times10^{-4}$&1.5565$\times10^{-4}$&1.3786$\times10^{-4}$\\
$\beta_4(ps^4/m)$&-5.9889$\times10^{-8}$&-1.4684$\times10^{-8}$&-1.2451$\times10^{-7}$\\
$\gamma_1 (W^{-1}m^{-1})$&0.64 &0.685&0.841 \\
$\gamma_2 (W^{-2}m^{-1})$&2.13$\times10^{-3}$&2.44$\times10^{-3}$&3.68$\times10^{-3}$\\
\hline
\end{tabular}
\end{table}
\begin{figure}
\centering
\subfigure[]{\label{normal24}\includegraphics[height=4.8 cm, width=6 cm]{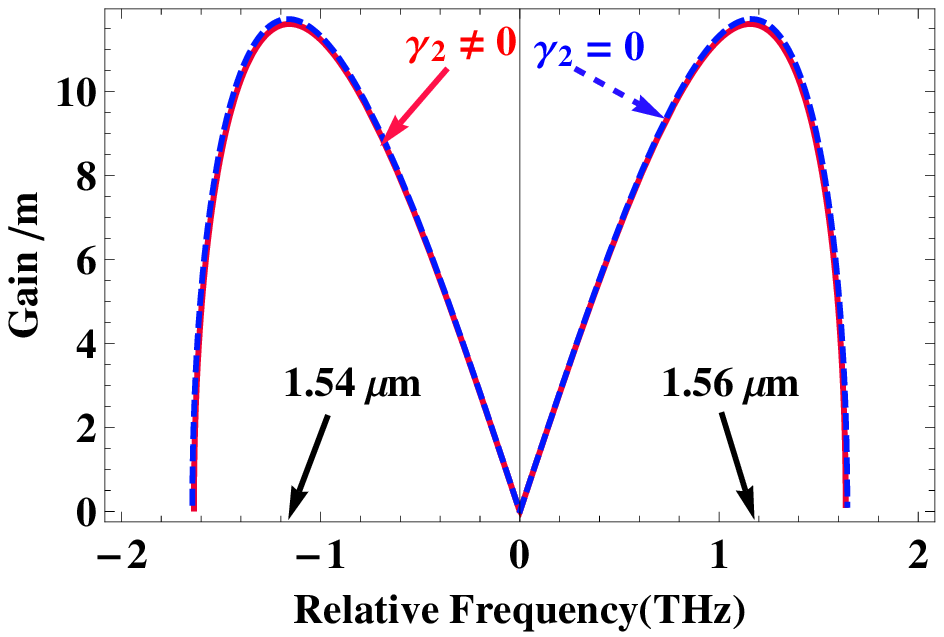}}
\subfigure[]{\label{normal24}\includegraphics[height=4.8 cm, width=6 cm]{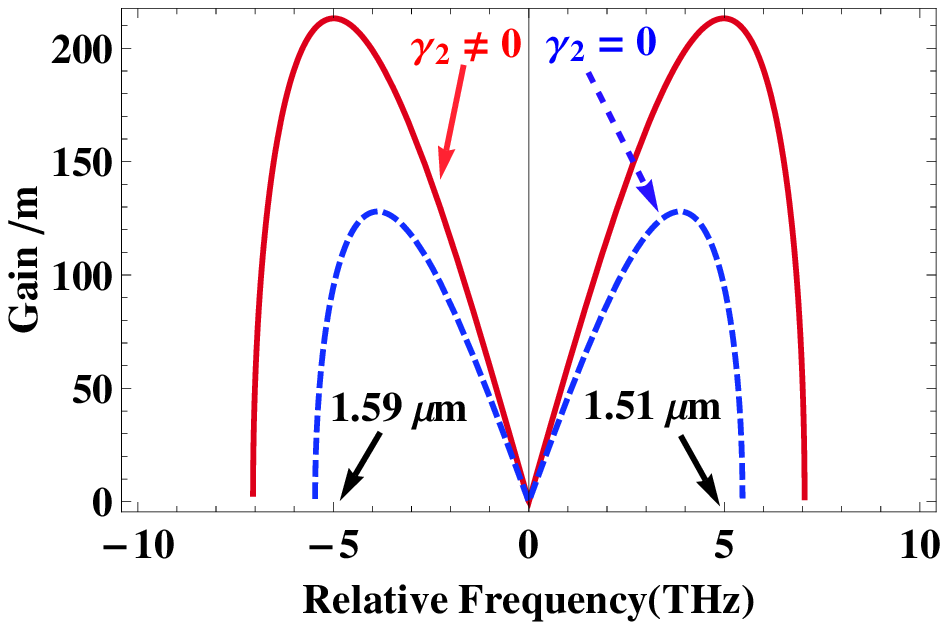}}
\caption{(Color online.) MI gain spectra of silica core (a) and $CS_2$ liquid core (b) PCF without suspension at P=100W.}
\label{MI_SF_0}
\end{figure}
\section{Modulation instability analysis}
 The phenomenon of modulational instability has been exploited a great deal to analyze the influence of higher-order effects on nonlinear processes such as soliton formation \cite{ Roo1, Hong, Hong2,  Roo2}, supercontinuum generation \cite{ak, vasa}, optical-switching \cite{ste} , etc. The formation of solitons and  MI analysis of X-ray laser beam in a relativistic quantum plasma has been investigated in \cite{ Roo1}. The MI of an extended nonlinear Schrodinger equation with third and fourth-order dispersions, cubic-quintic nonlinear terms, self steepening and intra-pulse Raman scattering describing the propagation of extremely short pulses have been investigated in Refs. \cite{Hong, Hong2}.  Also, the existence of MI and solitons in excitonic semiconductor waveguides has been studied \cite{Roo2}. The MI-induced supercontinuum generation in PCF using nanosecond pulses \cite{ak} and in the presence of saturable nonlinear response has been investigated in \cite{vasa}. The switching dynamics and MI of a nonlinear optical bistable system have been analyzed in Ref. \cite{ste}. Following this analysis, in the sections below, we perform an MI analysis to understand the impact of enhanced nonlinearity due to the infiltration of liquid and by the suspension on supercontinuum generation in the LCPCF.
\label{MI analysis}
\subsection{Propagation equation}
The equation governing the propagation of an optical beam in an optical fiber with higher-order dispersion and nonlinear effects is given by the modified nonlinear Schr\"{o}dinger equation (MNLSE) of the form as follows \cite{steq},
\begin{equation}
 i \frac{\partial Q(z,t)}{\partial z}- \frac{\beta_2}{2}\frac{\partial ^2Q(z,t)}{\partial t^2}- i\frac{\beta_3}{6}\frac{\partial ^3Q(z,t)}{\partial t^3}+ \frac{\beta_4}{24}\frac{\partial ^4Q(z,t)}{\partial t^4} \\ +\gamma_1|Q(z,t)|^2 Q(z,t)+\gamma_2|Q(z,t)|^4 Q(z,t)=0,
\label{Thm}
\end{equation}
where $Q(z,t)$ represents the electric field amplitude, $\beta_n$ is the dispersion coefficient of order $n$, where $n$ $=$ 2, 3 and 4, $\gamma_1$ and $\gamma_2$ are the cubic and quintic nonlinearity coefficients, respectively.
\subsection{Linear stability analysis}
\begin{figure}[ht]
\centering
\subfigure[]{\label{norm1}\includegraphics[height=4.7 cm, width=6 cm]{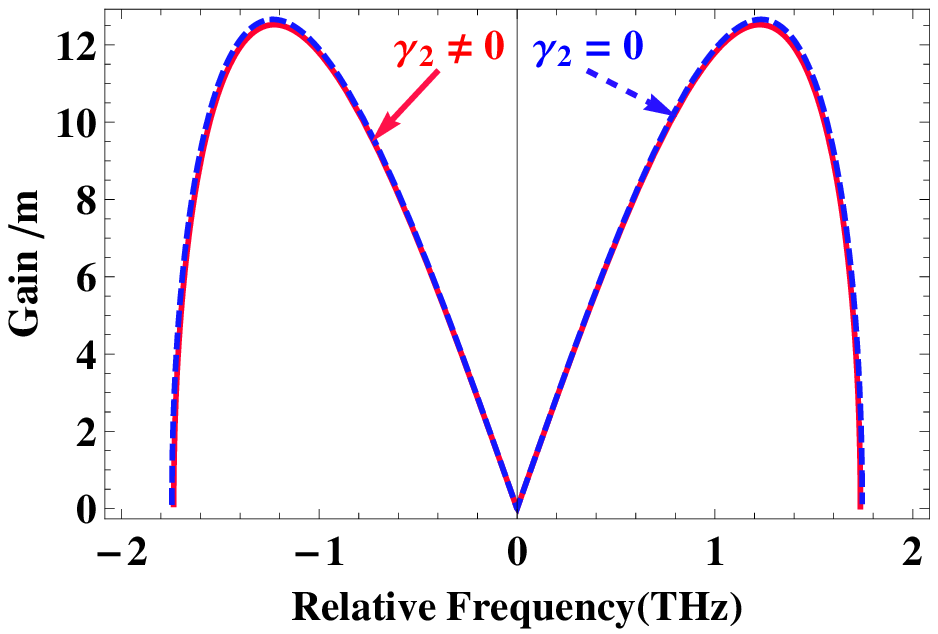}}
\subfigure[]{\label{norm2}\includegraphics[height=4.7 cm, width=6 cm]{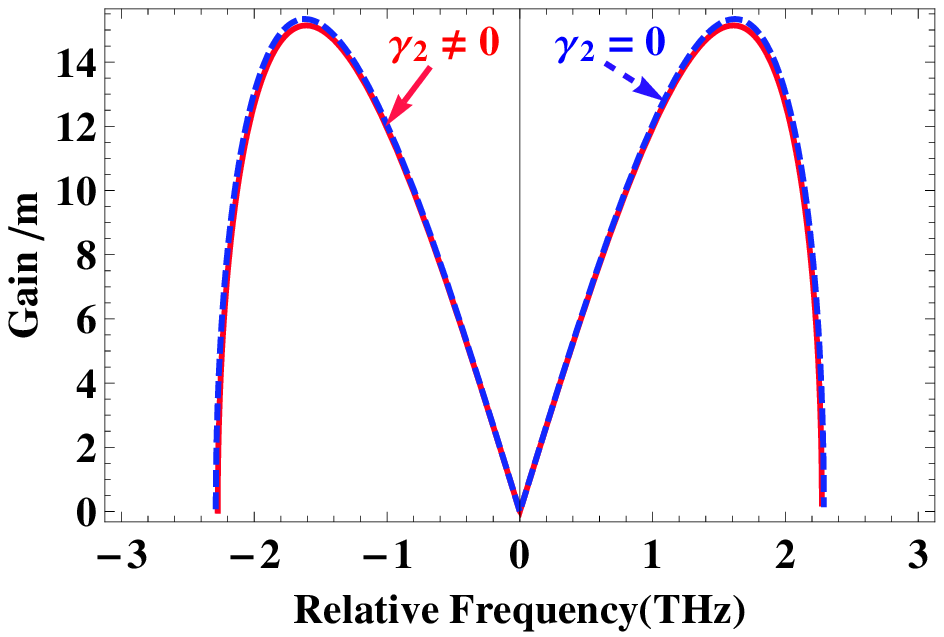}}
\subfigure[]{\label{norm3}\includegraphics[height=4.7 cm, width=6 cm]{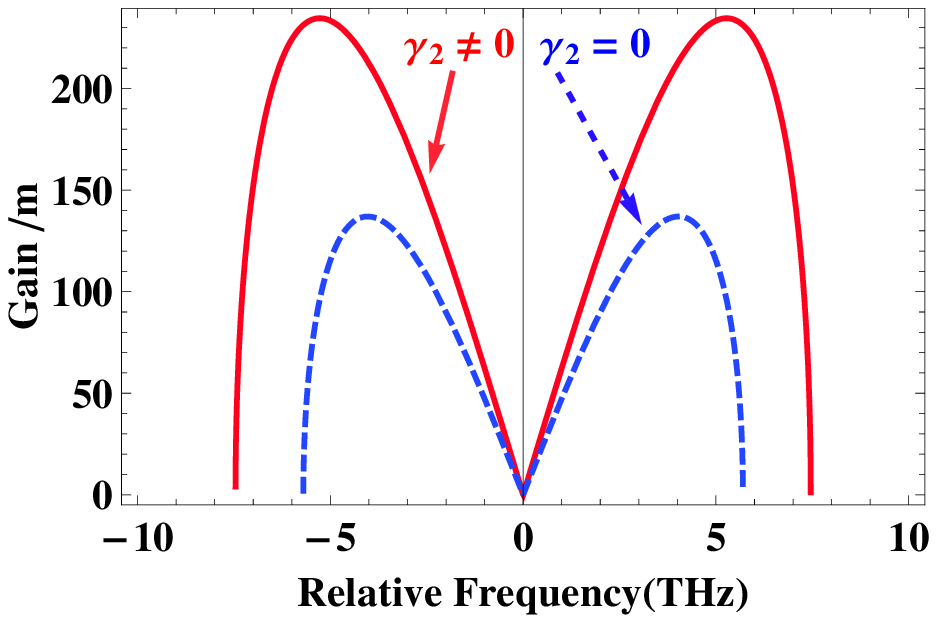}}
\subfigure[]{\label{norm4}\includegraphics[height=4.7 cm, width=6 cm]{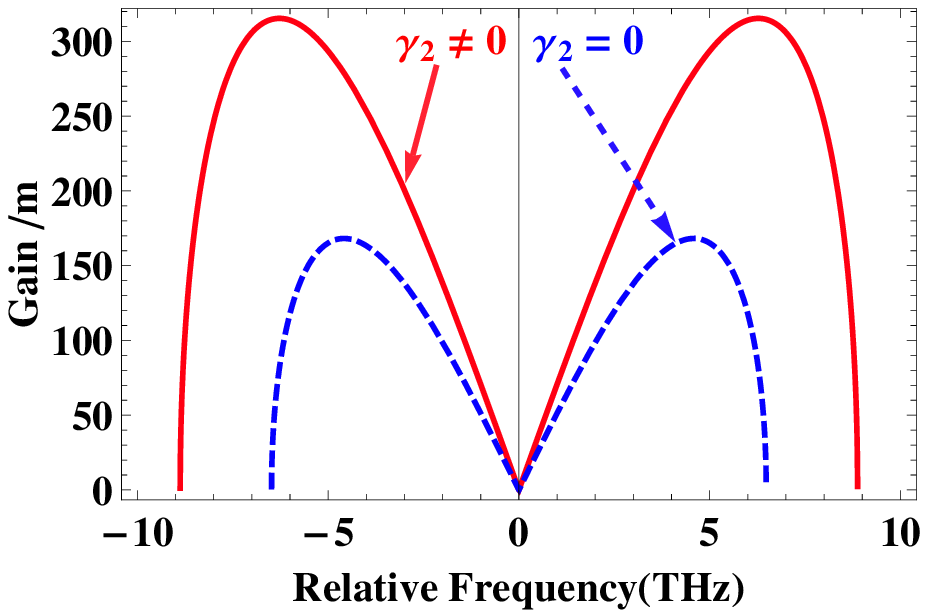}}
\caption{(Color online.) MI gain spectra of silica (first row) and $CS_2$ (second row) core  PCFs with (a and c) SF=1.222, (b and d) SF=1.444 at P=100W.}
\label{MI_SCPCF}
\end{figure}
The underlying framework of MI depends on the linear stability analysis where a plane wave solution is perturbed under the condition that the perturbation is very weak and then study whether the perturbation is growing or decaying during propagation. The stability of the steady-state solution against a weak perturbation in an optical fiber can be analyzed by utilizing the linear stability analysis (LSA).
The steady-state solution,\,$ i.e.$, the CW solution is given by,
\begin{equation}
\label{eq4}
Q(z,t)= \sqrt{P} \exp (i \Phi_{NL}).
\end{equation}
The dependence of the nonlinear phase shift $\Phi_{NL}$ with power ($P$) and propagation distance ($z$) is given by
\begin{equation}
\label{eq5}
\Phi_{NL}=P(\gamma_1+P \gamma_2)z.
\end{equation}
Now we introduce a very small perturbation to analyze the linear stability of the steady-state as
\begin{equation}
\label{eq6}
Q(z,t)= (\sqrt{P}+a(z,t)) \exp (i P(\gamma_1+P \gamma_2)z),
\end{equation}
where $|a(z,t)|<<\sqrt{P}$.\\
Substituting Eq. (\ref{eq6}) in Eq. (\ref{Thm}) we get a linearized equation,
\begin{eqnarray}
\label{eq6n}
\frac{\partial a(z,t)}{\partial z}+i \frac{\beta_2}{2}\frac{\partial ^2a(z,t)}{\partial t^2}-\frac{\beta_3}{6}\frac{\partial ^3a(z,t)}{\partial t^3}-i \frac{\beta_4}{24}\frac{\partial ^4a(z,t)}{\partial t^4}\nonumber\\-i P \gamma_1 a(z,t)^*-2i\gamma_2P^2 a(z,t)^*-iP\gamma_1a(z,t)-2i\gamma_2P^2 a(z,t)=0,
\end{eqnarray}
where $a(z,t)^*$ stands for the complex conjugate of $a(z,t)$.
In order to solve Eq. (\ref{eq6n}), we assume a plane wave ansatz consisting of two sideband components with forward and backward propagation having the form,
\begin{equation}
\label{ansatz}
a(z,t)=c [\exp{(i (K z-\omega t))}]+d [\exp{(-i (K z-\omega t))}],
\end{equation}
where $K$ and $\omega$ are the wave vector and frequency, respectively, of the perturbation amplitude. Substituting Eq.(\ref{ansatz}) in Eq.(\ref{eq6n}), we obtain a set of two linearly coupled equations satisfied by the coefficients $c$ and $d$, which is given by,
$$ \left(
   \begin{array}{cc}
     \epsilon_{11}&\epsilon_{12}\\
     \epsilon_{21}&\epsilon_{22}\\
   \end{array}
 \right)\left(
          \begin{array}{c}

            c\\
            d\\
          \end{array}
        \right)=0,
$$
 where $\epsilon_{11}=i K-\frac{i \beta_2}{2}\omega^2-\frac{i \beta_3}{6}\omega^3-\frac{i \beta_4}{24}\omega^4-i \gamma_1 P-2 i \gamma_2 P^2$, $\epsilon_{12}=-i \gamma_1 P-2 i \gamma_2 P^2$, $\epsilon_{21}=-i K-\frac{i \beta_2}{2}\omega^2+\frac{i \beta_3}{6}\omega^3-\frac{i \beta_4}{24}\omega^4-i \gamma_1 P-2 i \gamma_2 P^2$ and $\epsilon_{22}=-I \gamma_1 P-2 I \gamma_2 P^2$.
 This set has nontrivial solution only when the 2x2 determinant formed by the coefficient matrix vanishes. The vanishing condition on the determinant of the stability matrix gives the following dispersion relation \cite{Hong},
\begin{equation}
K=\frac{1}{6}\beta_3 \omega^3 |\omega|\pm\frac{i}{24}|\omega|\{-(\omega^2 \beta_4+12\beta_2)\times(\beta_4 \omega^4+12\beta_2\omega^2+48P \gamma_1+96P^2\gamma_2)\}^\frac{1}{2}.
\label{DispersionRelation}
\end{equation}
The above Eq. (\ref{DispersionRelation}) is the dispersion relation corresponding to the  LCSPCF. The MI gain can be derived from the above expression as $G=2 Im(K)$. The explicit expression for the gain of the LCSPCF can be written as,
\begin{equation}
\label{gain}
G=\frac{|\beta_4 \omega|}{12}[-(\omega^2+12\frac{\beta_2}{\beta_4})\times\{\omega^4+12\frac{\beta_2}{\beta_4} \omega^2+48\frac{P(\gamma_1+2 P \gamma_2)}{\beta_4}\}]^\frac{1}{2}.
\end{equation}
One can deduce the conventional and well known MI gain equation from Eq. (\ref{DispersionRelation}) by taking $\beta_3=\beta_4=\gamma_2=0$ with critical modulation frequency $\omega_c^2=2 P\gamma_1/|\beta_2|$. As the MI spectra bandwidth increases dramatically with an effective decrease in dispersion, the MI bandwidth becomes maximum in the vicinity of zero dispersion wavelength. It can also be observed that the MI gain increases with an increase in the nonlinearity coefficient of the fiber at the pump wavelength chosen. Hence for a better enhancement in MI, the fiber should have very low dispersion along with high nonlinearity coefficients at the pump wavelength considered. As described in the PCF design section of this paper, the proposed model possesses this fundamental property by increasing the suspension effect of the fiber core. Hence we expect an enhanced MI spectra corresponds to a suspended core fiber as compared to a conventional core fiber without suspension effect.
\section{Results and Discussion}
\label{Discussion}
Using the above expression for the MI gain (Eq. (\ref{gain})), in what follows, we briefly analyze the MI in the proposed LCSPCF under different parametric conditions of choice and compare the results with that of the conventional SCPCF. Emphasis will be given to the effect of core suspension and liquid infiltration in the instability gain expression. Using Eq. (\ref{gain}), we plot the instability spectra at the operating wavelength 1.55 $\mu m$ under different fiber settings given by the parameters described in table \ref{tab1}.  Figs. \ref{MI_SF_0}(a) and \ref{MI_SF_0}(b) represent the MI gain spectra  corresponding to the case of fiber without suspension (NS), and it is quite obvious that the LCPCF possesses high MI gain than the SCPCF due to the enhanced nonlinearity of the $CS_2$ infiltrated PCF. As the silica core PCF shows competing cubic-quintic nonlinearity ($\beta_2<0, \gamma_1>0$ and $\gamma_2<0$), the MI gain is suppressed with the effect of quintic nonlinearity.  The effect of quintic nonlinearity is almost negligible in our fiber setting for the case of SCPCF, while it shows a dramatic enhancement in MI gain and bandwidth in the case of LCPCF. This attribution is due to the cooperating cubic-quintic nonlinearity ($\beta_2<0, \gamma_1>0$ and $\gamma_2>0$)of $CS_2$ liquid resulting in an increased effective nonlinearity, thereby enhancing the gain of MI.

\begin{figure}[htb]
\centering
\subfigure[]{\includegraphics[height=4.8 cm, width=6 cm]{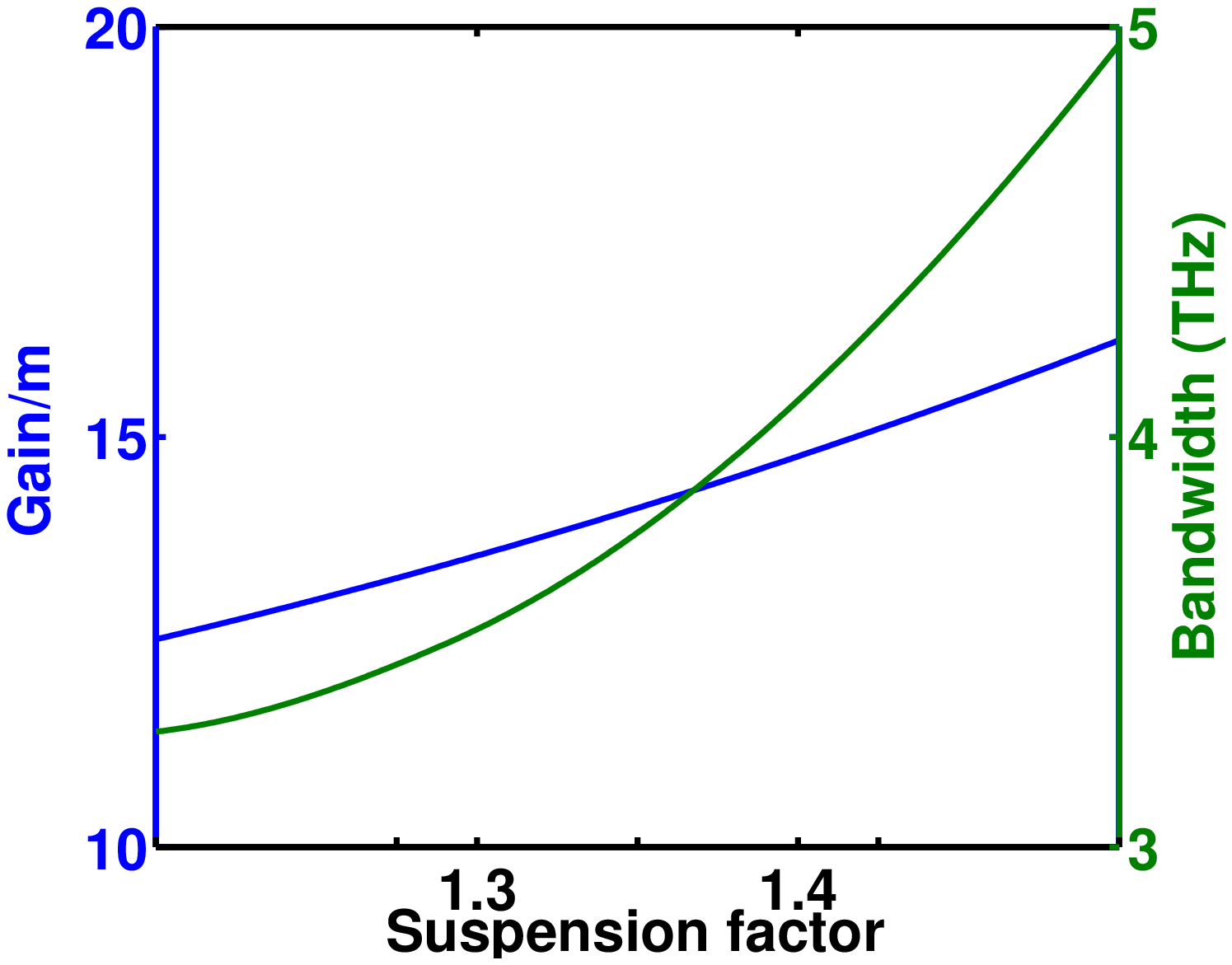}}
\subfigure[]{\includegraphics[height=4.8 cm, width=6 cm]{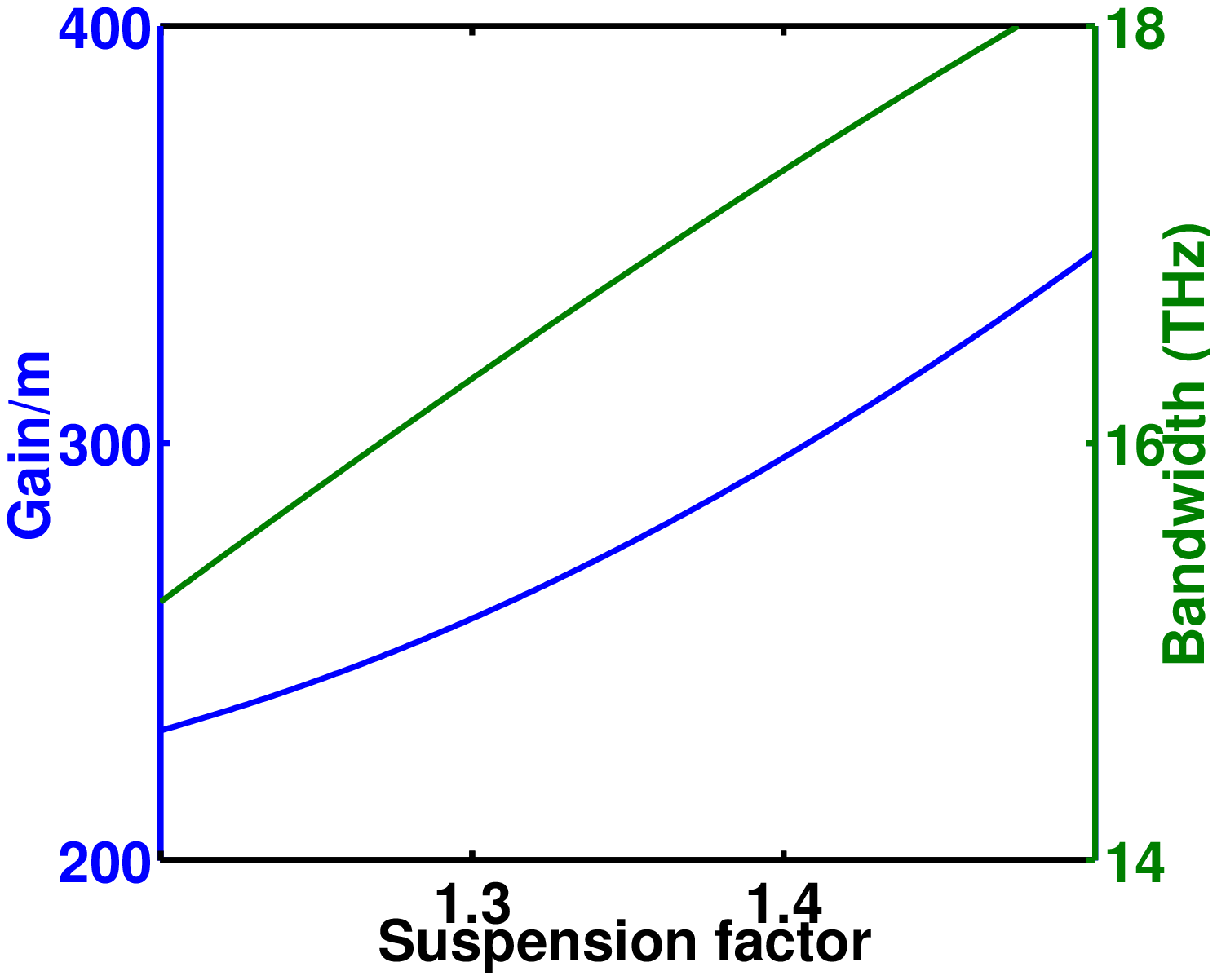}}
\caption{(Color online.) Variation of MI gain and bandwidth in silica PCF (a) and CS$_2$ (b) LCPCF for different suspension factors.}
\label{GAINBW}
\end{figure}

\begin{figure}[htb]
\centering
\subfigure[]{\includegraphics[height=6 cm, width=7 cm]{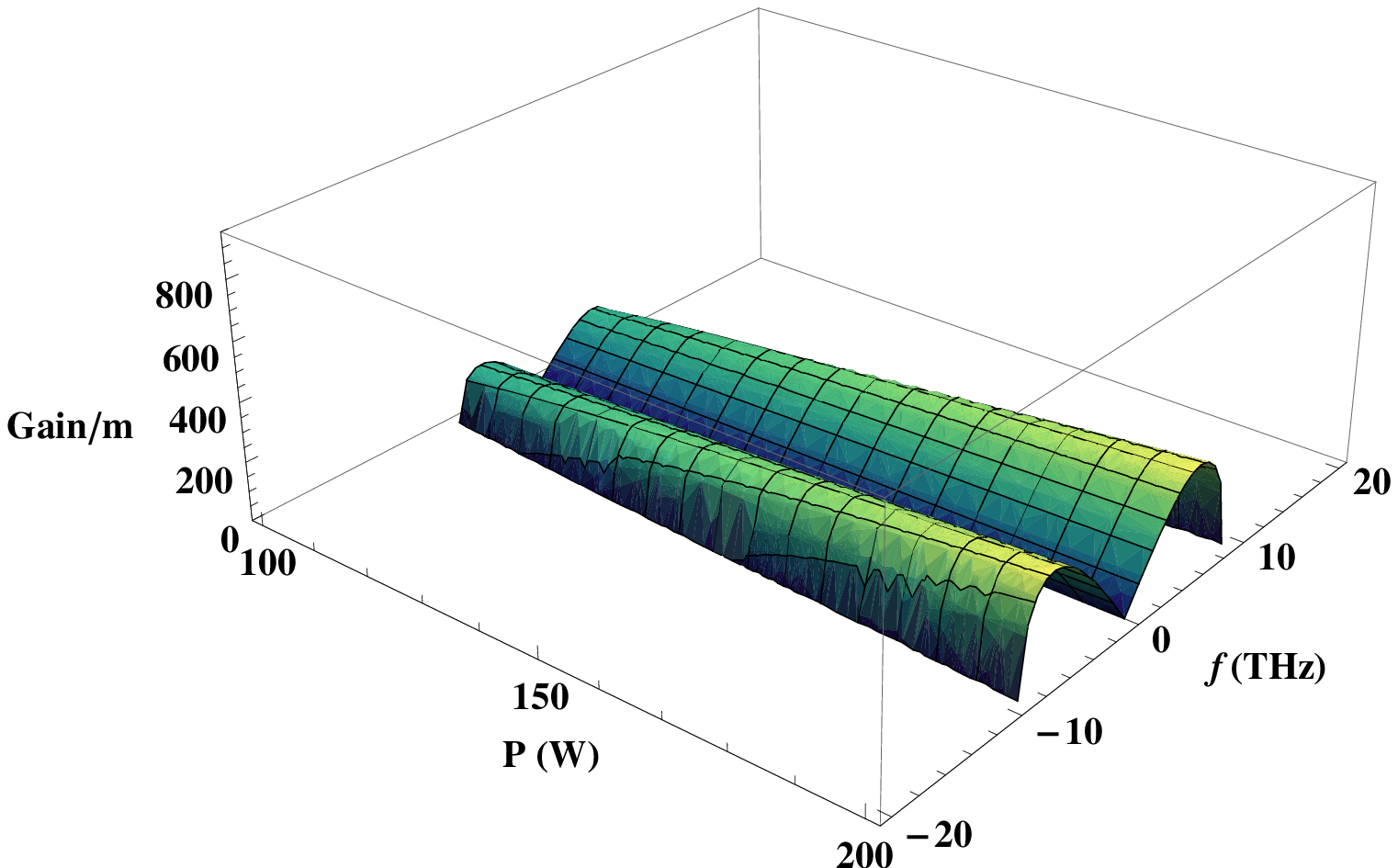}}
\subfigure[]{\includegraphics[height=6 cm, width=7 cm]{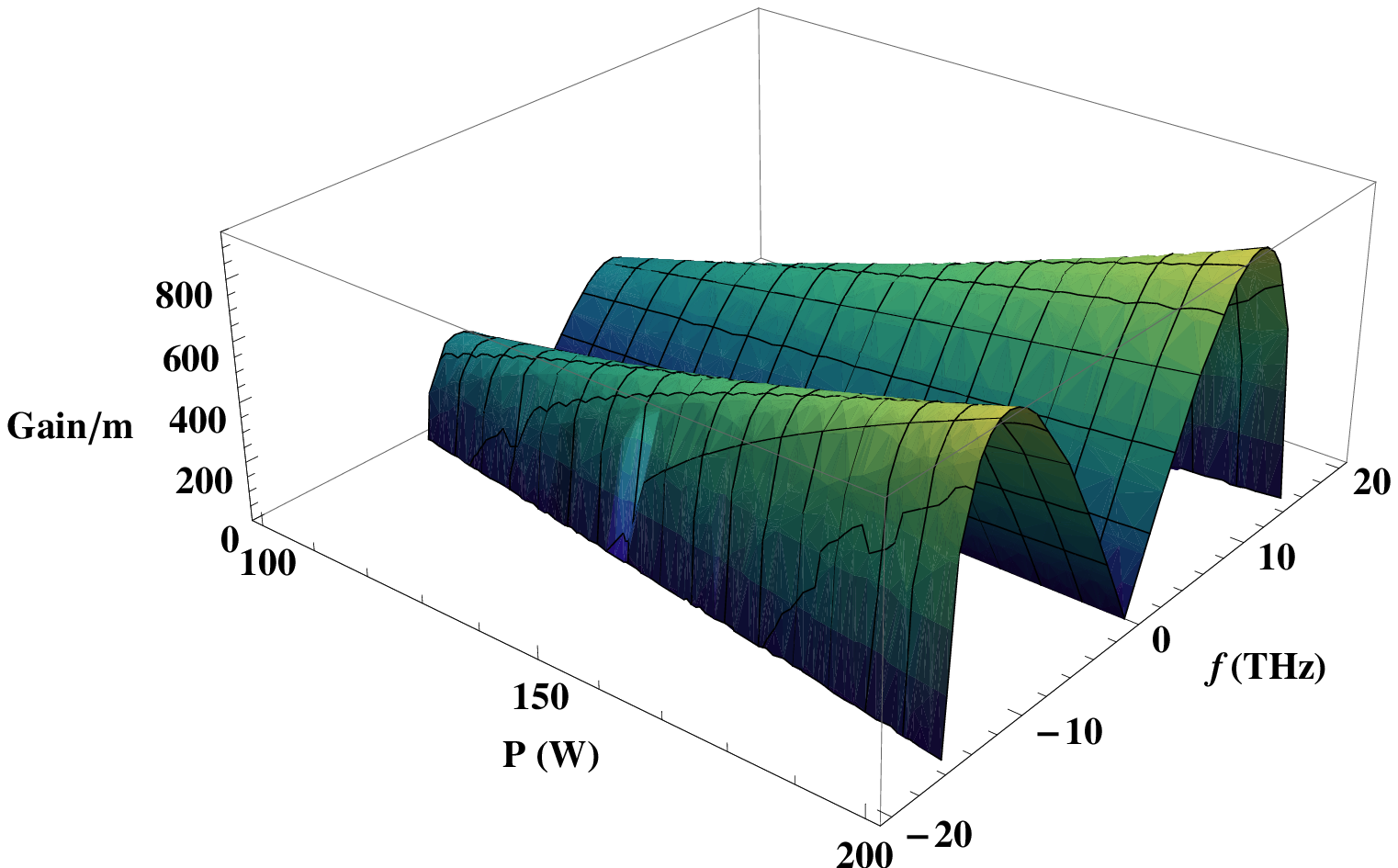}}
\caption{(Color online.) Variation of MI gain spectra for different input pump powers in LCSPCF (SF=1.444) without (a) and with (b) quintic nonlinearity.}
\label{MI_SCPCFP}
\end{figure}

Figs. \ref{MI_SCPCF} represents the instability spectra corresponding to SCSPCF and LCSPCF, respectively, for some representative values of suspension factor with and without quintic nonlinearity. Regardless of the fiber material in the core, it is quite evident from Fig. \ref{MI_SCPCF}  that the suspension inevitably increases the MI gain and bandwidth as a result of the increase in the effective nonlinearity due to core suspension.  Also one can notice that the effect of quintic nonlinearity is very minimal for SCSPCF, while it strongly affects the instability gain in the LCSPCF. For instance, the maximum gain produced by LCSPCF without quintic nonlinearity  which is 130$m^{-1}$ has been increased to 215$m^{-1}$ with the effect of quintic nonlinearity along with a considerable increase in spectral bandwidth. The shifts in maxima of the MI gain with and without quintic nonlinearity for $CS_2$ core LCPCFs are listed in table \ref{tab2}. For further insight, we plot in Fig. \ref{GAINBW}, the variation of MI gain and bandwidth in both silica and $CS_2$ core PCFs as a function of SF based on Eq. (\ref{gain}). The SF determines the effective area ($A_{eff}$) and effective index ($n_{eff}$) of the fundamental mode propagating through the core of the proposed PCF. The cubic nonlinearity, quintic nonlinearity and dispersion coefficients are the functions of $A_{eff}$ and $n_{eff}$ as given in Eqs. ((\ref{DispersionCoeff})-(\ref{QuinticNonlinearity})). Here we measure the values of $A_{eff}$ and $n_{eff}$ as a function of SF using COMSOL 5.2a and the variation of MI gain and bandwidth with respect to SF in both silica and $CS_2$ core PCFs are depicted in Fig. \ref{GAINBW}. It can be easily observed from the figure that the suspension increases the MI gain and bandwidth. This is due to the decrease in dispersion and increase in the nonlinear coefficients as the SF increases. It is also clear from Fig. \ref{GAINBW} that the enhancement of MI by increasing gain and bandwidth is high in the case of $CS_2$ core PCF as compared with silica core PCF. This is a result of enhancement of nonlinear effects in the PCF due to liquid infiltration. The effect of pump power in the instability spectra of LCSPCF without and with quintic nonlinearity is depicted in Figs. \ref{MI_SCPCFP}(a) and \ref{MI_SCPCFP}(b), respectively. We notice that irrespective of the nature of the fiber material and SF, the pump power increases the instability gain as it is the case of instability spectra reported in different contexts, whereas the rate of increase in gain for different pump power is larger when the quintic nonlinearity is present as compared to the case of cubic nonlinearity only. Thus the combined effect of infiltration and suspension leads to increased nonlinear coefficients owing to the high nonlinear refractive index of the liquid and the reduced effective area due to the core suspension. Also, as the ratio of air hole diameter to pitch (d/$\Lambda$) determines the effect of waveguide contribution in nonlinearity and dispersion, it is also possible to design fibers with various controllable properties. Also, it is evident from the phase-matching condition for MI, the maximum MI gain is achieved for the least value of dispersion coefficient, enabling an efficient interaction between the frequency components. Thus, the design of LCSPCF with the least $|\beta_2|$ would be much more desirable for different nonlinear applications. As the suspension effect causes a reduction in the value of the dispersion coefficient, it also plays a vital role in enhancing the MI spectra. Apart from the conventional method of reducing area by increasing the air hole diameter or by decreasing the pitch, the suspension core fiber enables a more flexible and desirable variation in dispersion and nonlinearity which enhances different nonlinear mechanisms like modulation instability and MI-induced SCG.
\begin{table}
\caption{\label{tab2} Maximum gain shift produced by silica PCF and $CS_2$ filled LCPCF with and without quintic nonlinearity.}
\begin{tabular}{|l|l|l|l|l|l|l|}
\hline
&\multicolumn{3}{l|}{Silica PCF}&\multicolumn{3}{l|}{CS$_2$ LCPCF}\\
\hline
Suspension& Maximum & Maximum & Gain shift &Maximum &Maximum &Gain shift\\
effect&gain without & gain with &\textbf{$(G_1-G_2)$}&gain without &gain with &\textbf{$(G_1-G_2)$}\\
 &$\, \gamma_2 (G_1)$ & $\,\gamma_2 (G_2)$& & $\, \gamma_2 (G_1)$& $\,\gamma_2 (G_2)$&\\
\hline
NS&11.71&11.61&0.1&130&215&85\\
SF=1.222&12.67&12.52&0.15&138&236&98\\
SF=1.444&15.36&15.12&0.24&171&317&146\\
\hline
\end{tabular}
\end{table}
\begin{figure}
\centering
\subfigure[]{\label{normal24}\includegraphics[height=4.5 cm, width=5.2 cm]{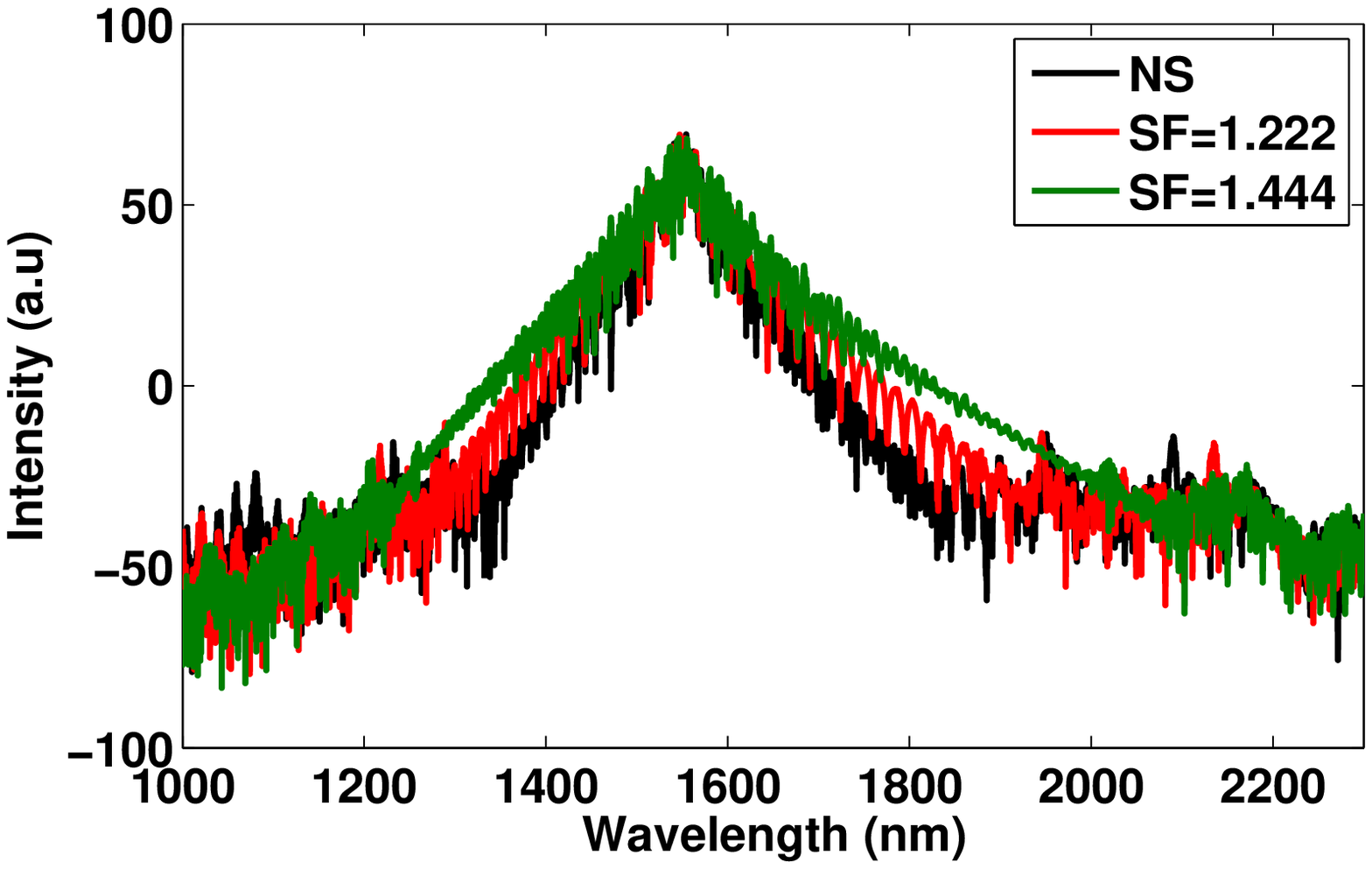}}
\subfigure[]{\label{normal24}\includegraphics[height=4.5 cm, width=5.2 cm]{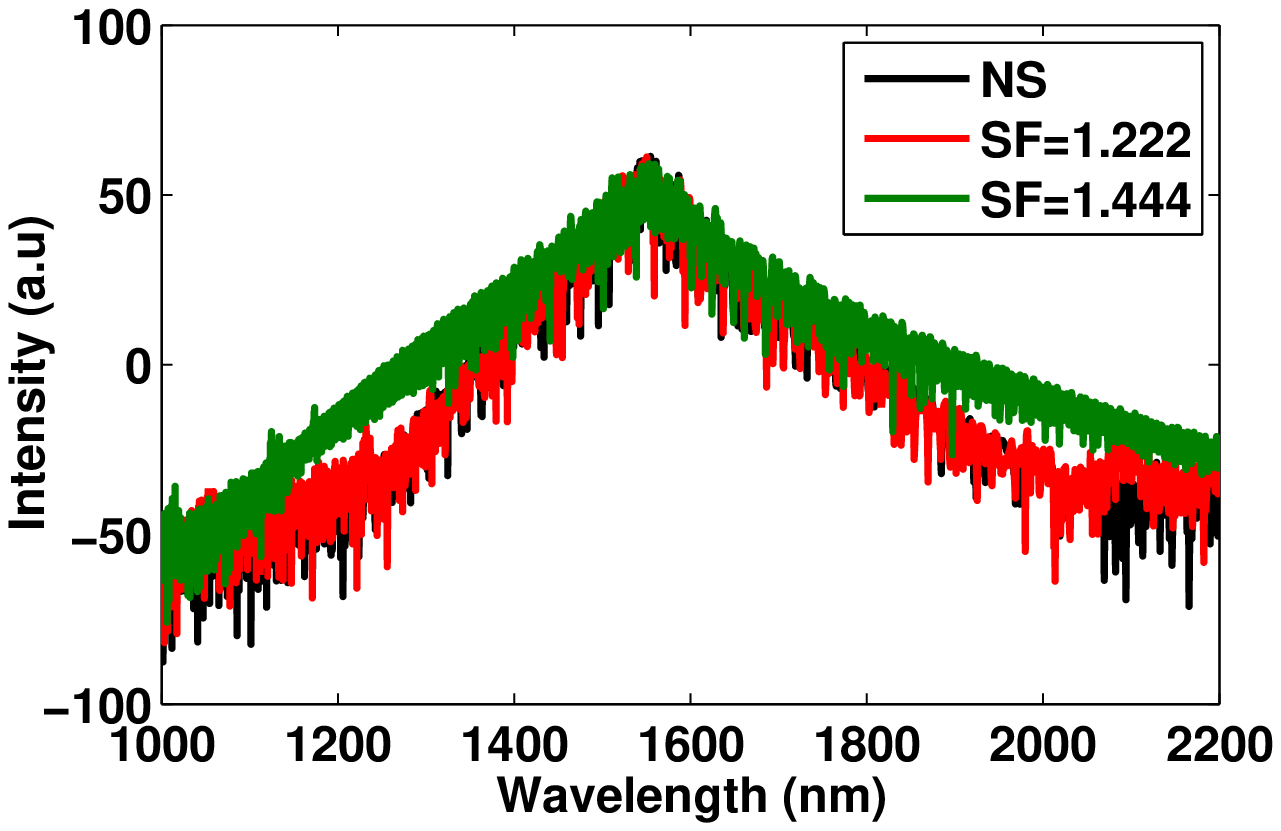}}
\caption{(Color online.) MI induced SCG in proposed (a) silica PCF (b) CS$_2$ LCPCF.}
\label{SCGBW}
\end{figure}
\section{Effect of suspension and quintic nonlinearity on MI induced SCG}
\label{MISCG}
In order to study the effect of suspension on the underlying nonlinear phenomena, we have considered the effect of SCG in all the proposed PCF models. We have used the equation governing the propagation of an ultrashort optical pulse in the PCF given by the modified nonlinear Schr$\ddot{o}$dinger equation described in Eq. (\ref{Thm}) for SCG studies. The MI-induced SCG is very similar to the soliton-driven SCG but the initial dynamics solely depends on the noise or perturbation of the input pulse. As the pulse propagates in the fiber, it gets modulated by the noise and also gets amplified exponentially.
This exponential increase in intensity leads to pulse braking, which further leads to spectral broadening. Since it is a noise-driven process the noise fraction in the spectra can be considerably larger for the MI-induced SCG process. The MI-induced SCG can be observed when the soliton fission length (L$_D$/N) is larger than the MI-induced pulse breaking distance. This can be easily achieved with the help of pulses with larger pulse widths as the dispersion length (L$_D$) is proportional to the pulse width (T) given by the expression L$_D$=T$^2$/$\beta_2$. Taking this into consideration, we have applied an input pulse of 5ps width for the SCG simulations.
\begin{figure}[htb]
\centering
\subfigure[]{\label{normal2421s}\includegraphics[height=5 cm, width=8 cm]{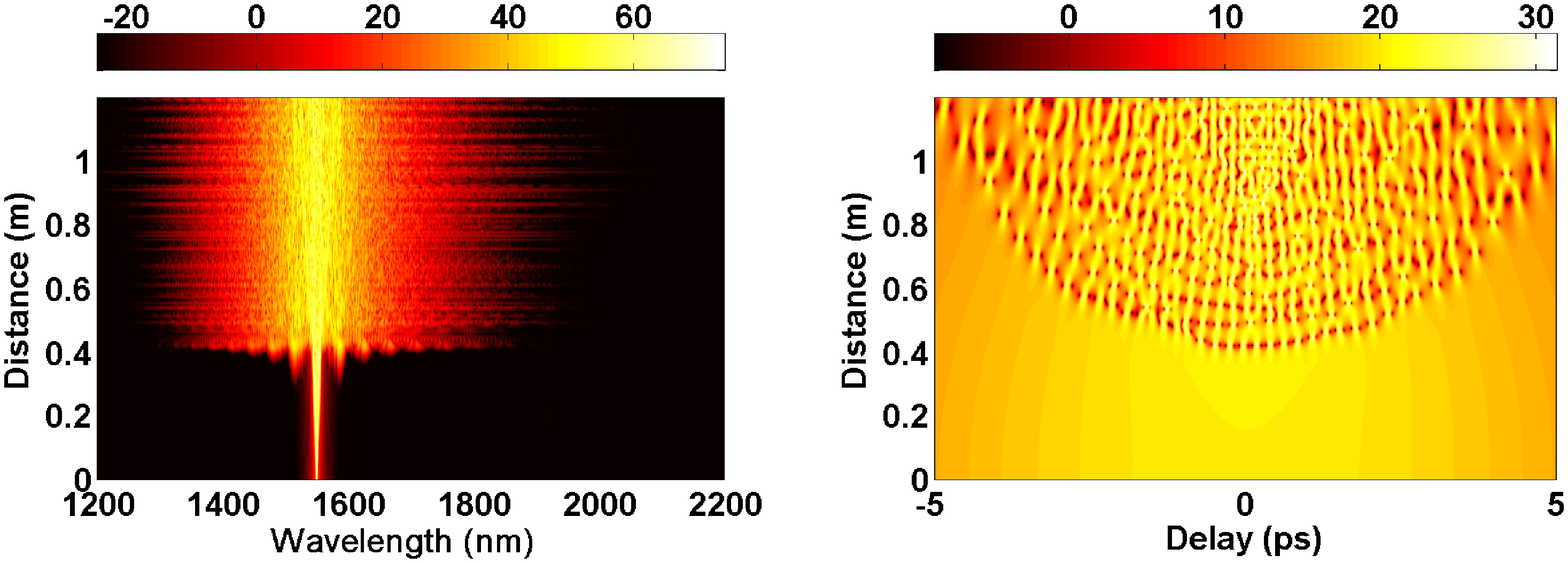}}
\subfigure[]{\label{normal24we}\includegraphics[height=5 cm, width=8 cm]{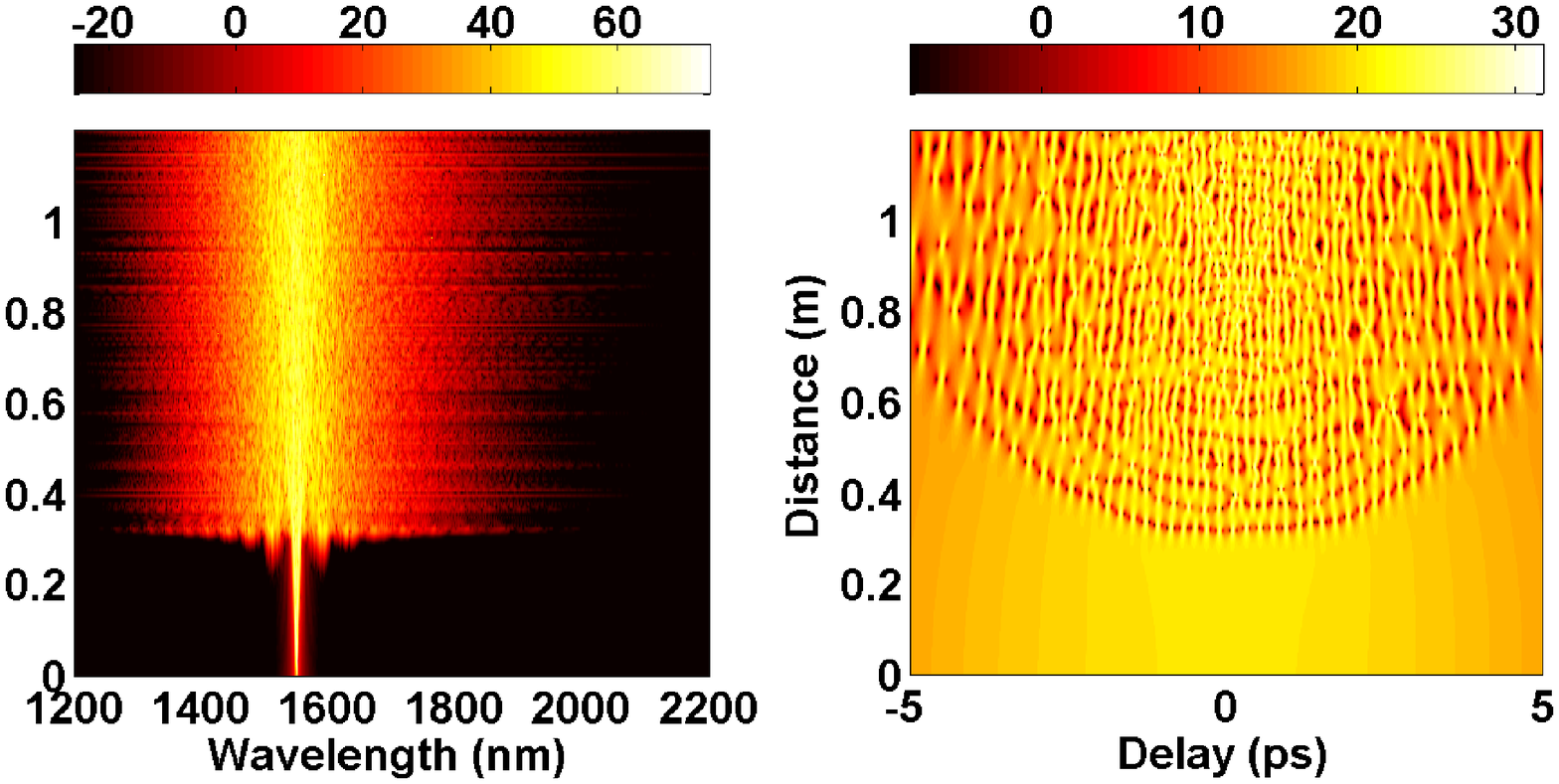}}
\caption{(Color online.) Spectral and temporal evolution in CS$_2$ LCPCF (a) without suspension and (b) with suspension factor, SF=1.44.}
\label{SCGBW12}
\end{figure}
\begin{figure}[htb]
\centering
\subfigure[]{\label{normal24}\includegraphics[height=5 cm, width=6 cm]{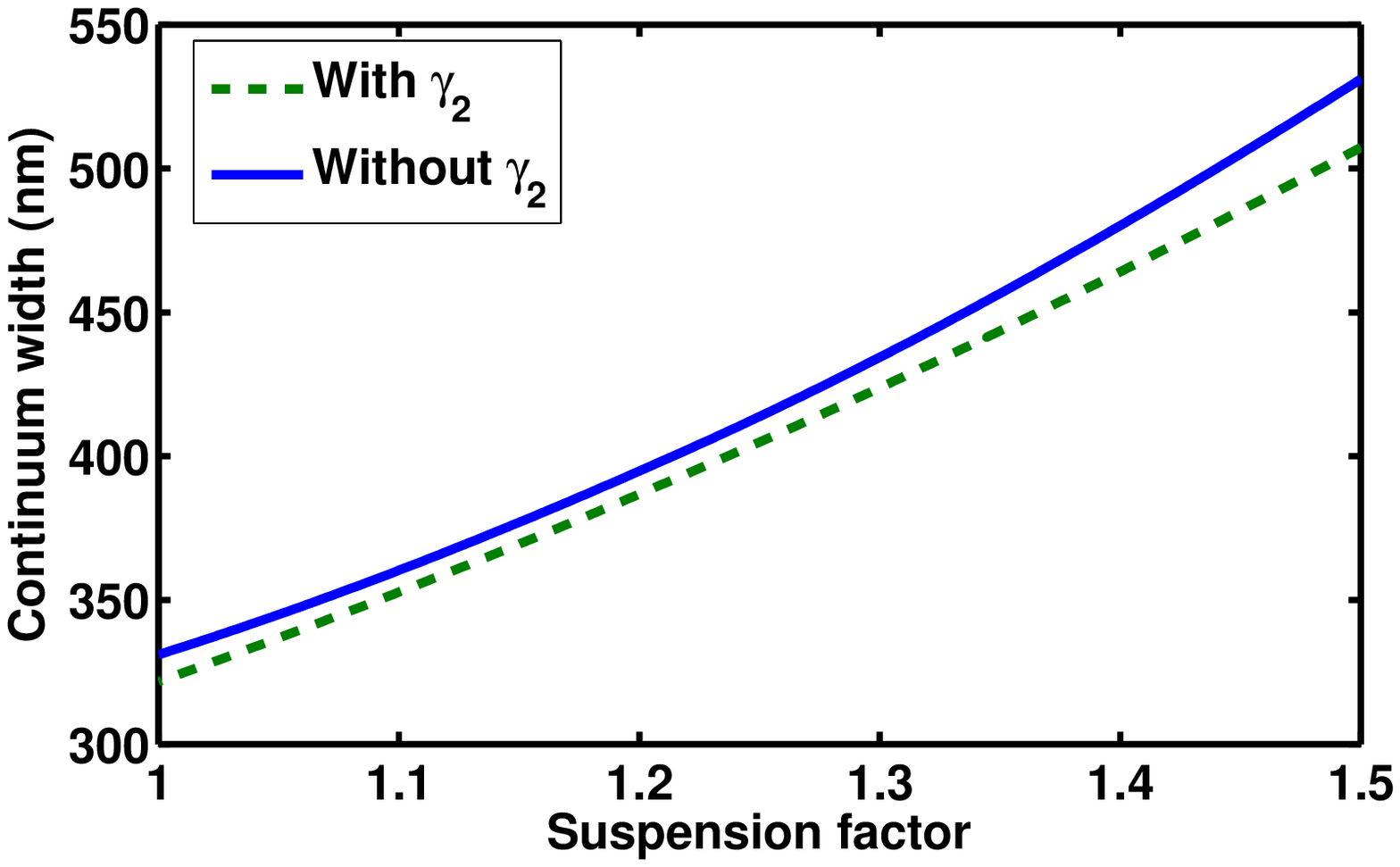}}
\subfigure[]{\label{normal24}\includegraphics[height=5 cm, width=6 cm]{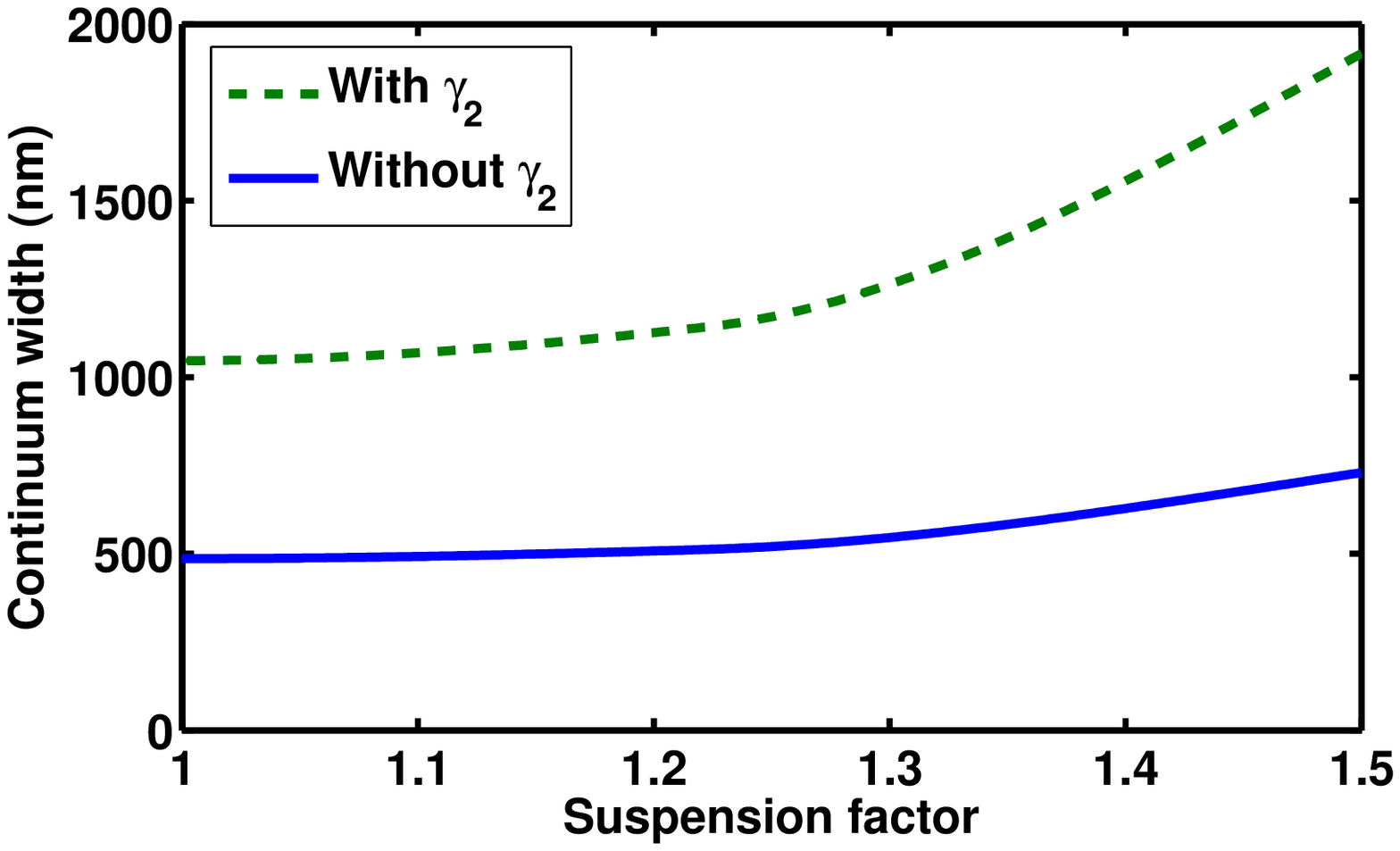}}
\caption{(Color online.) SC bandwidth variation with the effect of quintic nonlinearity for (a) silica PCF and (b) CS$_2$ LCPCF.}
\label{BWSF}
\end{figure}
\par
As the nonlinearity coefficients of the proposed PCF models increase along with a decrease in the dispersion coefficients, the MI induced supercontinuum width shows a consistent increase with the suspension effect. In our investigation, we initially study the effect of suspension on the SCG and hence the quintic nonlinearity term has been omitted from Eq. (\ref{Thm}) for this purpose. The MI-induced SCG in the proposed silica PCF and  CS$_2$ LCPCF are depicted in Figs. \ref{SCGBW}(a) and \ref{SCGBW}(b), respectively.  The spectral and temporal evolutions in CS$_2$ LCPCF without suspension are depicted in Fig. \ref{normal2421s}. The corresponding evolutions with suspension factor, SF=1.444, are shown in Fig. \ref{normal24we}.  All these figures show an unambiguous increase in the spectral bandwidth with the increase in the suspension factor. In order to perform a comparative study for each suspension factor, we have taken the minimum intensity for reference as 0 (a.u). When there is no suspension effect in silica PCF, the spectra spans a range of 331 nm. This has been further increased to 502 nm with suspension effect having SF=1.444 as shown in Fig. \ref{SCGBW}(a).
\par
\begin{table}
\caption{\label{tab3} Variation of SC bandwidth produced by silica PCF and $CS_2$ LCPCF with and without quintic nonlinearity.}
\begin{tabular}{|l|l|l|l|l|}
\hline
&\multicolumn{2}{l|}{Silica PCF}&\multicolumn{2}{l|}{CS$_2$ LCPCF}\\
\hline
Suspension&\,\,\,Bandwidth&Bandwidth&\,\,\,\,\,\,\,\,Bandwidth&Bandwidth\\
effect&\,\,\,without$\,\gamma_2 (nm)$ &with$\,\gamma_2$ (nm) &\,\,\,\,\,\,\,\, without$\,\gamma_2$ (nm) & with$\,\gamma_2$ (nm)\\
\hline
NS&\,\,\,331&321&\,\,\,\,\,\,\,\,486&1047\\
SF=1.222&\,\,\,403&394&\,\,\,\,\,\,\,\,512&1142\\
SF=1.444&\,\,\,502&482&\,\,\,\,\,\,\,\,672&1712\\
\hline
\end{tabular}
\end{table}
Further, we have studied the effect of quintic nonlinearity in the supercontinuum generation. Comparing the MI-induced SCG process with modulation instability gain and bandwidth, one can easily notice that the gain bandwidth product is directly proportional to the SC bandwidth. At first, we took the SCG produced by silica PCF with and without quintic nonlinearity as shown in Fig. \ref{BWSF}(a). From the figure one can easily understand that the spectral width has been decreased considerably with the effect of quintic nonlinearity. This can be easily linked with the decrease in gain and bandwidth of MI in the presence of quintic nonlinearity as shown in the MI spectra of silica PCF with different suspension factors. On the other hand, the effect of quintic nonlinearity is to further enhance the SC width in CS$_2$ LCPCF as the MI gain and bandwidth increase with the presence of quintic nonlinearity. The continuum width has increased from 672 nm to 1712 nm with the effect of quintic nonlinearity for LCSPCF with SF=1.444 as shown in Fig. \ref{BWSF}(b). For better insight, the variations of SC bandwidth produced by silica PCF and CS$_2$ LCPCF with and without quintic nonlinearity effect are tabulated in table \ref{tab3}. We have chosen the reference intensity as 0 (a.u) to measure the lower and upper cutoff wavelengths. The difference between the upper and lower cutoff wavelengths will give the bandwidth. From these observations, one can easily conclude that the cooperating nonlinearity has a positive effect on the MI-induced SCG whereas the competing nonlinearity affects the broadening negatively.
\section{SUMMARY AND CONCLUSION}
\label{conclusion}
In summary, we have introduced a liquid core suspended photonic crystal fiber as a potential contender for the next generation  of nonlinear optical devices. In this paper, we systematically studied the effect of liquid infiltration in the PCF and reported a dramatic enhancement of nonlinear effects. We proposed that the suspension of the PCF core would further enhance the nonlinearity due to the reduced effective core area. To facilitate a better understanding, a comparative illustration with the conventional solid-core PCF has been presented in all the cases studied throughout the manuscript. Thus we report that a combined effect of liquid infiltration and suspension significantly increases the nonlinear coefficient almost by two orders of magnitude. Also, in the proposed configuration, it is possible to control the dispersion by manoeuvring the SF and d/$\Lambda$ ratio, thus paving the way for delicate control on both the linear and nonlinear properties of the fiber.
\par
Following a detailed investigation on the various characteristics of LCSPCF, we demonstrated the functionality of the proposed LCSPCF in the enhancement of nonlinear processes by considering a fundamental nonlinear process, namely the MI. We have shown that the CS$_2$ LCSPCF dramatically increases the MI gain and bandwidth, owing to the elevated nonlinearity by the combined effect of liquid infiltration and suspension. Particularly, the effect of quintic nonlinearity in the instability spectra of SCPCF and LCSPCF is studied, and it is shown that the quintic nonlinearity has only a minimal effect in the solid core PCF, while it significantly increases the overall MI gain due to the cooperating effect of the dominant cubic and quintic nonlinearities in the CS$_2$ LCSPCF. Thus, we propose the CS$_2$ LCSPCF as a potential contender for the future generation of nonlinear applications such as MI, supercontinuum generation, pulse compression, etc. We have extended our study to incorporate the effect of quintic nonlinearity on MI-induced SCG in silica and CS$_2$ LCPCF. The cooperating nonlinearity present in the CS$_2$ LCPCF has a positive effect on MI-induced SCG which enables an enhancement in the SCG spectra whereas the competing nonlinearity which arises in silica PCF affects the broadening inversely. Although the suspended solid core PCF structures have been experimentally realized, we believe that the aforementioned theoretical insights will enable further studies, especially in the design and development of novel liquid filled suspended photonic crystal fibers for potential nonlinear applications.
\section{Acknowledgement}
 The author A.K.S. is supported by the University Grants Commission (UGC), Government of India, through a D. S. Kothari Post Doctoral Fellowship in Sciences. The work of M. L. is funded by DST-SERB through a Distinguished Fellowship (Grant No. SB/DF/04/2017).

\end{document}